\documentclass[12pt,preprint]{aastex}
\usepackage{natbib}
\usepackage{wrapfig}

\newcommand{\kms}{\ifmmode {\rm km~s}^{-1} \else km~s$^{-1}$\fi}
\newcommand{\ergs}{\ifmmode {\rm erg~ s}^{-1} \else erg~s$^{-1}$\fi}
\newcommand{\ergscm}{\ifmmode {\rm erg~s}^{-1} \else erg~s$^{-1}$ cm$^{-2}$\fi}
\newcommand{\Msun}{\ifmmode {\rm M}_{\odot} \else M$_{\odot}$\fi }
\newcommand{\Lsun}{\ifmmode {\rm L}_{\odot} \else L$_{\odot}$\fi}
\newcommand{\qo}{\ifmmode q_{\rm o} \else $q_{\rm o}$\fi}
\newcommand{\Ho}{\ifmmode H_{\rm o} \else $H_{\rm o}$\fi}
\newcommand{\ho}{\ifmmode h_{\rm o} \else $h_{\rm o}$\fi}

\newcommand{\vFWHM}{\ifmmode v_{\mbox{\tiny FWHM}} \else
                    $v_{\mbox{\tiny FWHM}}$\fi}
\newcommand{\CCF}{\ifmmode F_{\it CCF} \else $F_{\it CCF}$\fi}
\newcommand{\ACF}{\ifmmode F_{\it ACF} \else $F_{\it ACF}$\fi}
\newcommand{\Halpha}{\ifmmode {\rm H}\alpha \else H$\alpha$\fi}
\newcommand{\Hbeta}{\ifmmode {\rm H}\beta \else H$\beta$\fi}
\newcommand{\Hgamma}{\ifmmode {\rm H}\gamma \else H$\gamma$\fi}
\newcommand{\Hdelta}{\ifmmode {\rm H}\delta \else H$\delta$\fi}
\newcommand{\Lya}{\ifmmode {\rm Ly}\alpha \else Ly$\alpha$\fi}
\newcommand{\Lyb}{\ifmmode {\rm Ly}\beta \else Ly$\beta$\fi}
\newcommand{\HeI}{\ifmmode {\rm He}\,{\sc i}\,\lambda5876 \else 
	          He\,{\sc i}\,$\lambda5876$\fi}
\newcommand{\HeII}{\ifmmode {\rm He}\,{\sc ii}\,\lambda4686 \else 
	           He\,{\sc ii}\,$\lambda4686$\fi}

\newcommand{\hii}{H\,{\sc ii}}

\newcommand{\ciii}{\ifmmode {\rm C}\,{\sc iii} \else C\,{\sc iii}\fi}

\newcommand{\oi}{O\,{\sc i}}
\newcommand{\oii}{O\,{\sc ii}}

\newcommand{\mbh}{$M_{\rm BH}$\ }

\shorttitle{Nuclear X-ray sources in SINGS Galaxies}
\shortauthors{Grier et al.}

\begin{document}

\title{Discovery of Nuclear X-ray Sources in SINGS Galaxies}

\author{C.~J.~Grier\altaffilmark{1},
S.~Mathur\altaffilmark{1},
H.~Ghosh\altaffilmark{2},
L.~Ferrarese\altaffilmark{3}}

\altaffiltext{1}{Department of Astronomy, The Ohio State University,
140 W 18th Ave, Columbus, OH 43210; grier@astronomy.ohio-state.edu} 
\altaffiltext{2}{CNRS/CEA-Saclay, 91911 Gif-sur-Yvette, France}
\altaffiltext{3}{Hertzberg Institute of Astrophysics, 5071 West 
Saanich Road, Victoria, BC, V9E 2E7, Canada}

\begin{abstract}
We present the results of a search for nuclear X-ray activity in
nearby galaxies using $Chandra$ archival data in a sample of 62
galaxies from the $Spitzer$ Infrared Nearby Galaxy Survey (SINGS). We
detect 37 nuclear X-ray sources; seven of these are new
detections. Most of the nuclear X-ray sources are likely to be
AGNs. The fraction of galaxies hosting AGNs is thus about 60\%, much
higher than that found with optical searches, and demonstrates the
efficacy of X-ray observations to find hidden AGNs in optically normal
galaxies. We find that the nuclear X-ray sources are preferentially
present in earlier type galaxies. Unlike what is observed at high
redshift for high-luminosity AGNs, we do not find a strong correlation
between the AGN luminosity and the $24\mu$m luminosity of the host
galaxy; we find a strong correlation with the $3.6\mu$m luminosity
instead. This suggests that at the present epoch the accretion rate
depends on the total mass of the galaxy, as perhaps does the black
hole mass.
\end{abstract}

\section{INTRODUCTION}
  
The past decade has seen extraordinary growth in our understanding of
supermassive black holes (SMBHs), with secure detections, mass
measurements and new demographic information (see
\citealt{Ferrarese05} and references therein; FF05
hereafter). Knowledge of the mass function of SMBHs directly affects
our understanding of SMBH formation and growth, nuclear activity, and
the relation of SMBHs to the formation and evolution of galaxies in
hierarchical cold dark matter models (e.g. \citealt{Menci04}). The
cumulative mass function needed to explain the energetics of high
redshift quasars implies that all galaxies in the local universe
should host a SMBH (e.g. \citealt{Marconi04}, \citealt{Shankar04}), but
we do not know whether this is indeed the case.

Many SMBHs have been found, and their masses measured through stellar
and gas dynamics methods at the high end of the SMBH mass function
(FF05). In the intermediate mass range, around 10$^8$ \Msun, we know
of the existence of SMBHs as active galactic nuclei (AGNs); their
masses are measured through reverberation mapping or using scaling
relations with emission line widths and luminosity
(e.g. \citealt{Blandford82}, \citealt{Peterson93}, \citealt{Wandel99},
\citealt{Kaspi00}, 2005). When considering smaller SMBHs or galaxies
farther away, mass measurement is more difficult; the sphere of
influence of the black hole in these objects is too small to resolve,
so we cannot use dynamical estimates, and if we do not see AGN
activity at visible wavelengths, we cannot use reverberation
mapping. For these reasons, the low-mass end of the SMBH mass function
is largely unexplored; indeed we still do not know whether all
galaxies host SMBHs.

Because of these difficulties, we turn to AGN signatures to help us
probe the existence of low-mass SMBHs. The observed ``downsizing'' of
AGNs (e.g. \citealt{Hasinger05}), in which more luminous AGNs are seen
to be accreting more actively at higher redshifts and less luminous
objects are expected to dominate in the local universe, implies that
proportionally more low-mass SMBHs should still be accreting at
present. The low mass SMBHs are expected to be present in late-type
galaxies. It is possible that the nearby ``normal'' galaxies appear
``normal'' because AGN activity is weak and is washed out by starlight
when viewed in the optical bands. However, if these objects are in
fact accreting, we should observe many of them in X-ray
emission. X-rays are well-suited to identifying low-mass systems in
late-type galaxies because they can penetrate obscuring material,
which in many cases can mask out the emission-line regions surrounding
the AGN. Additionally, for low luminosity AGNs the optical light can
be overwhemed by the host galaxy light; X-rays, once again, are
helpful in this respect because galaxies themselves are not very bright in
X-rays. Indeed, there have been several successful efforts using
X-rays to identify AGN activity in galaxies residing in clusters
(\citealt{Martini02}) and in fields (\citealt{Brand05}).

In an effort to improve our knowledge of the low-mass end of the SMBH
mass function, we conducted a search for low-luminosity AGN candidates
using $Chandra$ 
(\citealt{Ghosh08}, 2009, 2010a, 2010b). Our program was successful in
that we discovered AGNs in what were thought to be normal
galaxies. Through extensive spectral, timing and multiwavelength
analysis we classified nuclear X-ray sources in 56 galaxies and found
17 that are almost certainly AGN. Thus at least 30\% of normal
galaxies are actually active. The inferred luminosities of these
sources range from $10^{37.5}$ to $10^{42}$ erg s$^{-1}$. In a few
objects where SMBH masses were known from stellar/gas dynamics
methods, we find accretion rates as low as $10^{-5}$ of the Eddington
limit.

We found that AGNs are present in galaxies of all Hubble types. The
distribution of luminosities in a given type and across Hubble types
is wide. Moreover, in a given Hubble type, some galaxies host AGNs
(down to our flux limit), but some do not. Thus the important
outstanding question is: What governs accretion onto a SMBH? At low
redshift, the merger rate is low, so the accretion is unlikely
to be merger driven. Is it related to star formation rate, as seen at
high redshift (e.g. \citealt{Netzer09b}, \citealt{Lutz08})? Is it the
galactic structure in the central regions, such as bars or nuclear
spirals (\citealt{Pogge02})? Multiwavelength data are necessary to
answer these fundamental questions and the $Spitzer$ Infrared Nearby
Galaxy Survey (SINGS) provides such data. Most SINGS galaxies have been
observed with $Chandra$, so we searched for nuclear X-ray sources in
these galaxies.  The main goals of this study were to investigate in
detail the X-ray properties of the SINGS sample, detect potential new
low-luminosity nuclear activity, and investigate possible connections
between nuclear X-ray luminosity and various multiwavelength
properties of the host galaxies.

\section{DATA AND ANALYSIS}
\subsection{Galaxy Sample}
Our galaxy sample consists of objects from SINGS and is described in
detail in \cite{Kennicutt03}. The original sample contained 75
galaxies within 30 Mpc that were chosen to obtain a broad range of
galaxy properties to represent the composition of the local
universe. This sample was well-suited for our study because these
galaxies have been the subjects of many different investigations at
many different wavelengths, and we hoped to investigate various
properties of galaxies and their connection to the nuclear X-ray
activity. Basic information on each of the galaxies in our sample was
retrieved from \cite{Kennicutt03} and presented in Table
\ref{Table:tbl1} along with information on the $Chandra$ observations
used. There were 13 objects in the SINGS sample for which data was
unavailable in the $Chandra$ archive; the remaining 62 SINGS galaxies
had been observed with $Chandra$ by the time of this study and the
data are available for public use. These 62 galaxies constitute the
sample in our study. The SINGS sample is composed of both quiescient
and active galaxies; our X-ray sample includes 13 objects whose nuclei
are optically identified as Seyferts, nine that are identified as \hii
\ regions, 11 that contain low ionization nuclear emission regions
(LINERs), and seven that are optically identified as starbursts. The
remaining 22 galaxies in our study had no listed optical nuclear
classifications.

\subsection{Data Analysis}
We used the $Chandra$ Interactive Analysis of Observations software
(CIAO)\footnote{http://cxc.harvard.edu/ciao3.4/index.html} v3.4 to
process images and extract source counts
(\citealt{Fruscione06}). X-ray data were downloaded from the $Chandra$
archive and filtered to exclude background flares, following the
procedure of \cite{Ghosh08}. We used the CIAO task $wavdetect$ to
determine the positions of the sources. We then searched for nuclear
sources using object positions listed in the NASA$/$IPAC Extragalactic
Database (NED). $Chandra$'s 90\% source location error circle has a
radius of less than $0''\!.6$ and less than 1\% of sources fall
outside a 1\arcsec \ radius. Therefore, sources that were detected
within a 1\arcsec \ radius of the listed position in NED are
considered nuclear detections.

Source counts were extracted from a region centered on the source
found by $wavdetect$; the radius of the source region is equal to the
95\% encircled-energy radius or $2''\!.3$ (4.67 pixels), whichever is
greater. We extracted the background counts from an annulus with an
outer radius of 5 times the radius of the source region and an inner
radius of twice the radius of the source region. Any other source
that fell inside this annulus was excised from the image before
counts were extracted.  Broadband source counts were extracted in the
0.3-8.0 keV range; counts were also extracted in the 0.3-2.5 keV
(soft) and the 2.5-8.0 keV (hard) bands. We also calculated the
hardness ratios, defined as HR = ($H - S$)/($H + S$), where H
represents the number of hard counts and S represents soft counts,
when possible. 

We retrieved $Spitzer$ IRAC, MIPS, and 2MASS fluxes from
\cite{Dale07}, which are measured from image mosaics that are large
enough to detect emission out to R$_{25}$ in each galaxy. These flux
densities therefore represent the total fluxes of the galaxy at a
given wavelength. We also searched for 2MASS point source detections
at the positions of the nuclei in these galaxies; all of our detected
X-ray sources have 2MASS counterparts that are listed in the 2MASS
point source catalog (\citealt{Skrutskie06}). The 2MASS magnitudes are
measured using a PSF profile-fitting algorithm (see catalog for
details) within an aperture of $\sim$4\arcsec \ centered on the position
of the detected source. We also obtained 1.4 GHz integrated flux
densities for those sources that were detected in the VLA ``Faint
Images of the Radio Sky at Twenty-Centimeters'' survey (FIRST;
\citealt{White97}). The FIRST survey has a resolution of about
5\arcsec; fluxes were measured by fitting a Gaussian to the
observations. To summarize, the X-ray, 2MASS source, and FIRST
luminosities are nuclear luminosities, while the $Spitzer$ and 2MASS
global luminosities from \cite{Dale07} include light from the entire
galaxy.

\section{X-RAY RESULTS}

\subsection{New Detections}
We detected nuclear sources in 37 objects in this sample. Six of these
detections are in galaxies that have not yet been examined for nuclear
X-ray activity using $Chandra$ observations. We also report a new
detection in one galaxy (NGC\,855) that has been previously searched
but its nucleus had not been detected. We discuss the new detections
individually below. All quoted X-ray luminosities are calculated using
a power law with $\Gamma = 2$ and Galactic absorption.

In order to quantify our assumptions for AGN activity, we computed IR
to X-ray slopes in these new objects to see how they compared with
observed slopes in known AGNs. We compare our measured slopes with
\cite{Laor94}, who measure the average IR to soft X-ray slope
$<\alpha_{\rm{irs}}>$ = $-1.26$ $\pm$ 0.11 in their sample of quasars,
defining $\alpha_{\rm{irs}}$ as
\begin{equation}
\alpha_{\rm{irs}} = \rm{log} (f_{0.3 \rm{keV}}/f_{1.69 \mu\rm m}) / 2.611.
\end{equation}
We estimate $\alpha_{\rm{irs}}$ for the seven objects discussed
individually below. We calculate $f_{0.3 \rm{keV}}$ by using a
$\Gamma$ = 2 power law and our observed integrated 0.3-8 keV fluxes to
estimate the flux at 0.3 keV. As mentioned, we have 3.6 $\mu$m fluxes
for all of these objects from \cite{Dale07}, which we use to
approximate $f_{1.69 \mu\rm m}$. Spectral templates from
\cite{Assef2010} show that for a typical AGN, the difference in flux
between 1.69 $\mu$m and 3.6 $\mu$m is about a factor of $1/3$. We
correct our fluxes by this factor, which translates to an increase in
$\alpha_{\rm{irs}}$ in our sources of 0.18. Typical uncertainties in
$\alpha_{\rm{irs}}$ for these 7 measurements are around 0.2. We can
then directly compare our slopes with the results of Laor et al;
individual measurements are discussed below.

\subsubsection{NGC\,1404}
\begin{figure}
\begin{center}
\epsscale{0.48}
\plotone{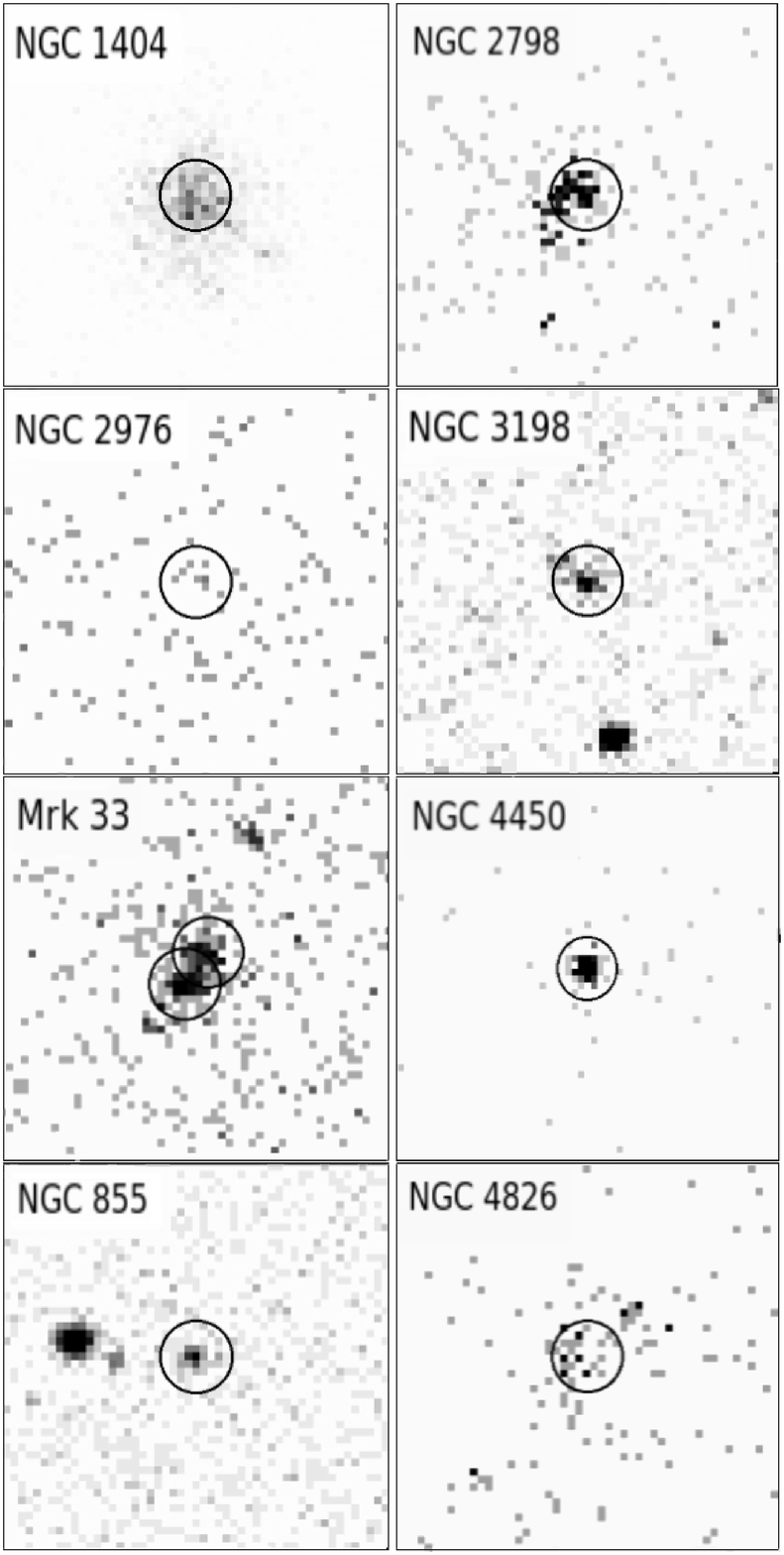}
\caption{Broad band (0.3-8 keV) $Chandra$ images for the eight sources
with new X-ray detections. Each image is 25\arcsec \ $\times$
25\arcsec \ in size.}
\label{fig:f1}
\end{center}
\end{figure}

NGC\,1404 is a type E1 galaxy at a distance of 25.1 Mpc. There was no
previously listed nuclear classification for this object. There are
several publications investigating NGC\,1404 in X-rays but to our
knowledge, none of them examine the nuclear X-ray properties of the
galaxy. We have four observations of NGC\,1404 and it was detected in
all four; we here report on only the longest observation. NGC\,1404
was observed by $Chandra$ using the ACIS-I camera for $\sim$44 ks and
detected with 797 net broadband counts. Its hardness ratio
HR$=-0.904$. We estimate $L_{0.3-8 \rm keV}$ = 1.55 $\times$ 10$^{40}$
\ergs \ for this source. This point source was detected in the 2MASS
survey with L$_J$ = 7.6 $\times$10$^{43}$ \ergs, L$_H$ = 1.4
$\times$10$^{44}$ \ergs, and L$_K$ = 2.9 $\times$10$^{44}$
\ergs. Given the 2MASS luminosities, the bolometric luminosity of the
source is high enough for it to be an AGN. We measure
$\alpha_{\rm{irs}}$ = $-1.35$; this slope is consistent with that
expected of an AGN as measured by \cite{Laor94}. Based on the
$Chandra$ image (Figure~\ref{fig:f1}), there is a potential for
extended emission as well, which could originate from the
circumnuclear region, as observed in Seyfert 2 galaxies
(\citealt{Ghosh07}).

\subsubsection{NGC\,2798}
This is an SBa galaxy at a distance of 24.7 Mpc that was observed for
$\sim$5 ks. Its nucleus was listed by \cite{Kennicutt03} and
references therein as a starburst region based on optical
spectroscopy. We detect the nucleus with 73 counts; the nuclear source
is shown in Figure \ref{fig:f1}. Its hardness ratio is HR = $-0.82$
and we estimate $L_{0.3-8 \rm keV}$= 7 $\times$ 10$^{39}$ \ergs. The
nucleus was detected by 2MASS with observed $L_J$ = 4.8
$\times$10$^{43}$ \ergs, $L_H$ = 1.1 $\times$10$^{44}$ \ergs, and
$L_K$ = 1.7 $\times$10$^{44}$ \ergs.  This source is also detected in
FIRST with an integrated 1.4GHz flux density of 61.38 mJy. We measure
$\alpha_{\rm{irs}}$ = $-1.16$, again consistent with the slope
expected in AGNs. Because this nuclear source is also an X-ray and
radio source and has such high 2MASS luminosities, we identify it here
as an AGN, though there is likely a starburst contribution to the
X-ray flux.

\subsubsection{NGC\,2976} 
This SAc galaxy is at a distance of 3.5 Mpc, and its nucleus is listed
as an \hii \ region in \cite{Kennicutt03}. It was observed for
$\sim$10 ks and the nucleus was detected with 6 broadband counts. The
source is shown in Figure~\ref{fig:f1}.  There are insufficient counts
to recover a spectrum and HR, but we estimate $L_{0.3-8 \rm keV}$= 6.2
$\times$ 10$^{36}$ \ergs, again using a standard power law model. The
nucleus was detected by 2MASS with $L_J$ = 3.5 $\times$10$^{40}$
\ergs, $L_H$ = 7.6 $\times$10$^{40}$ \ergs, and $L_K$ = 1.0
$\times$10$^{41}$ \ergs. The 2MASS luminosities of this source are not
as high as the two galaxies listed above, but the $K-$band luminosity
is high enough that the source is unlikely to be an X-ray binary. We
measure $\alpha_{\rm{irs}}$ = $-1.9$, which is somewhat steeper than
expected, but is consistent with an AGN with an obscured X-ray source.

\subsubsection{NGC\,3198} 
NGC\,3198 is an SBc galaxy at a distance of 9.8 Mpc. It was observed
by $Chandra$ for 61.8 ks and was detected with 108 broadband counts,
corresponding to log $L_{0.3-8 \rm keV}$= 1.4 $\times$ 10$^{38}$
\ergs. The source is shown in Figure~\ref{fig:f1}. The X-ray hardness
ratio HR = $-0.535$.  The nucleus was detected by 2MASS with $L_J$ =
1.2~$\times$10$^{42}$ \ergs, $L_H$ = 2.4~$\times$10$^{42}$ \ergs, and
$L_K$ = 4.0~$\times$10$^{42}$ \ergs. The nucleus is identified in the
FIRST survey with an integrated flux density of 1.89 mJy. We measure
$\alpha_{\rm{irs}}$ = $-1.65$, which is again consistent with the
presence of an AGN with an obscured central X-ray source. As in the
case of NGC\,2798 above, this is highly likely to be an AGN based on
its 2MASS luminosities and the fact that it is both an X-ray and radio
source.

\subsubsection{Mrk 33}
Mrk 33 is categorized as an irregular galaxy in NED but is listed by
SIMBAD as a pair of interacting galaxies at a distance of 21.7
Mpc. Its nucleus is also listed as a starburst region by
\cite{Kennicutt03}. The nucleus was detected with two separate
components. One source is detected in both the soft and hard bands,
but the other source disappears completely in the hard band. Here we
focus only on the hard source; while a soft-only source could be an
AGN if it is completely obscured and all we see is
reprocessed/starburst emission, we cannot be sure without a good
quality spectrum. Because the 95\% energy radii of the two sources
overlap, we were unable to disentangle the two sources
completely. Since the counts in the soft band are similar in each
source, we extracted counts from a region encircling both sources and
halved the number of soft counts to obtain counts for the hard
source. We attribute all of the hard counts to the hard source. Our
X-ray image is shown in Figure \ref{fig:f1}. The harder source is
detected with 100 broadband counts, corresponding to a luminosity
$L_{0.3-8 \rm keV}$= 2.2 $\times$ 10$^{39}$ \ergs. The hardness ratio
is HR= $-0.640$. The hard source was detected by 2MASS with $L_J$ =
1.3 $\times$10$^{43}$ \ergs, $L_H$ = 2.2 $\times$10$^{43}$ \ergs, and
$L_K$ = 2.8 $\times$10$^{43}$ \ergs. It is also a radio source,
detected in the FIRST survey with an integrated flux density of 9.56
mJy. Again, its high 2MASS luminosities and that it is both an X-ray
and a radio source merit its inclusion as an AGN
candidate. Furthermore, we measure $\alpha_{\rm{irs}}$ = $-1.08$ in
this source, again is consistent with our expectations in AGNs. As
noted above, we cannot be sure that the soft source is an
AGN. However, it is reasonable to suspect that it too is an obscured
AGN at the center of the second galaxy of the pair. Much follow up is
required on this particular object to determine if this is in fact the
case.

\subsubsection{NGC\,4450} 
NGC\,4450 is an Sab galaxy at a distance of 20 Mpc and its nucleus has
been optically identified as a LINER. It was observed by $Chandra$ for
3.7 ks and detected with 223 counts. We estimate $L_{0.3-8 \rm keV}$=
5.0 $\times$10$^{40}$ \ergs \ and hardness ratio HR = $-0.695$. The
source is shown in Figure \ref{fig:f1}. Its nucleus was detected by
2MASS with $L_J$ = 3.8 $\times$10$^{43}$ \ergs, $L_H$~=~7.6
$\times$10$^{43}$ \ergs, and $L_K$ = 8.1~$\times$10$^{43}$ \ergs. We
measure $\alpha_{\rm{irs}}$ = $-1.04$ in this source, which again is
consistent with expectations in AGNs to within our uncertainties. It
is also a radio source detected by FIRST with an integrated flux
density of 6.53 mJy, and is clearly an AGN.

\subsubsection{NGC\,855} 
There has been some disagreement regarding the morphological
classification of NGC \,855. While this object is listed as an
elliptical, many authors in fact argue that this is a dwarf elliptical
galaxy. We discuss its classification further in section 5. We have
one 59.48 ks observation of the NGC\,855 and detect the nucleus with
108 broad band counts. \cite{Zhang09} used a 1.7 ks exposure of this
object and did not detect the source-- this makes our detection a new
detection. The nuclear source is shown in Figure~\ref{fig:f1}. We
estimate $L_{0.3-8 \rm keV}$ = 1.6 $\times$10$^{38}$ \ergs. The
nucleus was also detected by 2MASS with $L_J$ = 1.5 $\times$10$^{42}$
\ergs, $L_H$ = 1.5~$\times$10$^{42}$ \ergs, and $L_K$ = 1.9~
$\times$10$^{42}$ \ergs.  We measure $\alpha_{\rm{irs}}$ = $-1.35$ in
this source, which again is consistent with expectations in AGNs; this
object is very likely an AGN.

\subsection{Detections in objects with previous publications}
There were 44 objects on which previous analysis has been done;
previous publications that investigate the nuclei of these galaxies
using $Chandra$ observations are listed in Table \ref{Table:tbl1}. We
analyzed these as well to provide a consistent analysis of the entire
sample. Our measurements are presented in Tables \ref{Table:tbl2} and
\ref{Table:tbl3}. With the exception of NGC\,4826, all of our results
are consistent with previous measurements. Our measurements for this
object are presented below. In the cases of NGC\,3521, NGC\,3627,
NGC\,4569, NGC\,4594, and NGC\,4725, we have analyzed data with longer
exposure times than have previously been published.

\subsubsection{NGC\,4826} 
NGC\,4826 is an SAab galaxy that is optically identified as a Seyfert 2
(\citealt{Veroncetty06}). We detect a point source here, while neither
\cite{Ho01} or \cite{Zhang09} report a nuclear point source
detection. Figure~\ref{fig:f1} shows our $Chandra$ image. The detected
source is soft, and therefore unlikely to be direct AGN emission.
However, this could be an obscured AGN in which only circumnuclear
emission is visible, as is often the case with Seyfert 2 galaxies 
(\citealt{Ghosh07}).


\subsection{Non-detections}
We do not detect a nuclear source in 25 of the galaxies in our
sample. Several of these objects are irregular galaxies; for these
objects we searched for X-ray sources within a few arcseconds of the
galaxy position as listed on NED, since in these cases the ``center''
of the galaxy is poorly defined. 3$\sigma$ upper limits to all
non-detections are given in Table \ref{Table:tbl2}; the upper limits
were calculated by extracting source counts within a $2''\!.3$ radius
circle centered on the nuclear position as listed on NED and
background counts from an annulus with the same measurements as the
background regions for the detected sources. In all cases, our results
are consistent with previous results. In the cases of Ho IX,
NGC\,4625, NGC 4826, and NGC\,5474, we used data with longer exposure
times than previous studies, but we still do not detect the nuclei in
these sources.

\subsection{$Chandra$ Results: Summary}
Altogether, out of the 75 SINGS galaxies, 62 have data in the
$Chandra$ archive and we detect nuclear X-ray sources in 37 of
them. The nuclear X-ray sources, however, could be stars, binaries,
supernova remnants or AGNs. In the Ghost et al. papers (\S 1) we did
extensive spectral, timing and multiwavelength analysis to identify
the nuclear X-ray sources. Similarly, we show that the new detections
in SINGS galaxies listed above are highly likely to be AGNs by
examining their properties in multiple wavelengths as well as
comparing their IR to X-ray slopes with those of identified AGNs. As
discussed in \cite{Ghosh10a}, when the sample as a whole is considered
it becomes statistically unlikely that all detected nuclear X-ray
sources are contaminants which happen to be at the center of the
galaxy (see also \citealt{Ho09}, \citealt{Zhang09}). We therefore
argue that statistically, most, if not all, of the nuclear X-ray
sources in SINGS galaxies are AGNs; we assume them to be AGNs for the
rest of this paper.

\section{MULTIWAVELENGTH ANALYSIS}

\subsection{$Spitzer$ Measurements}
\begin{figure}
\begin{center}
\epsscale{1.0}
\plotone{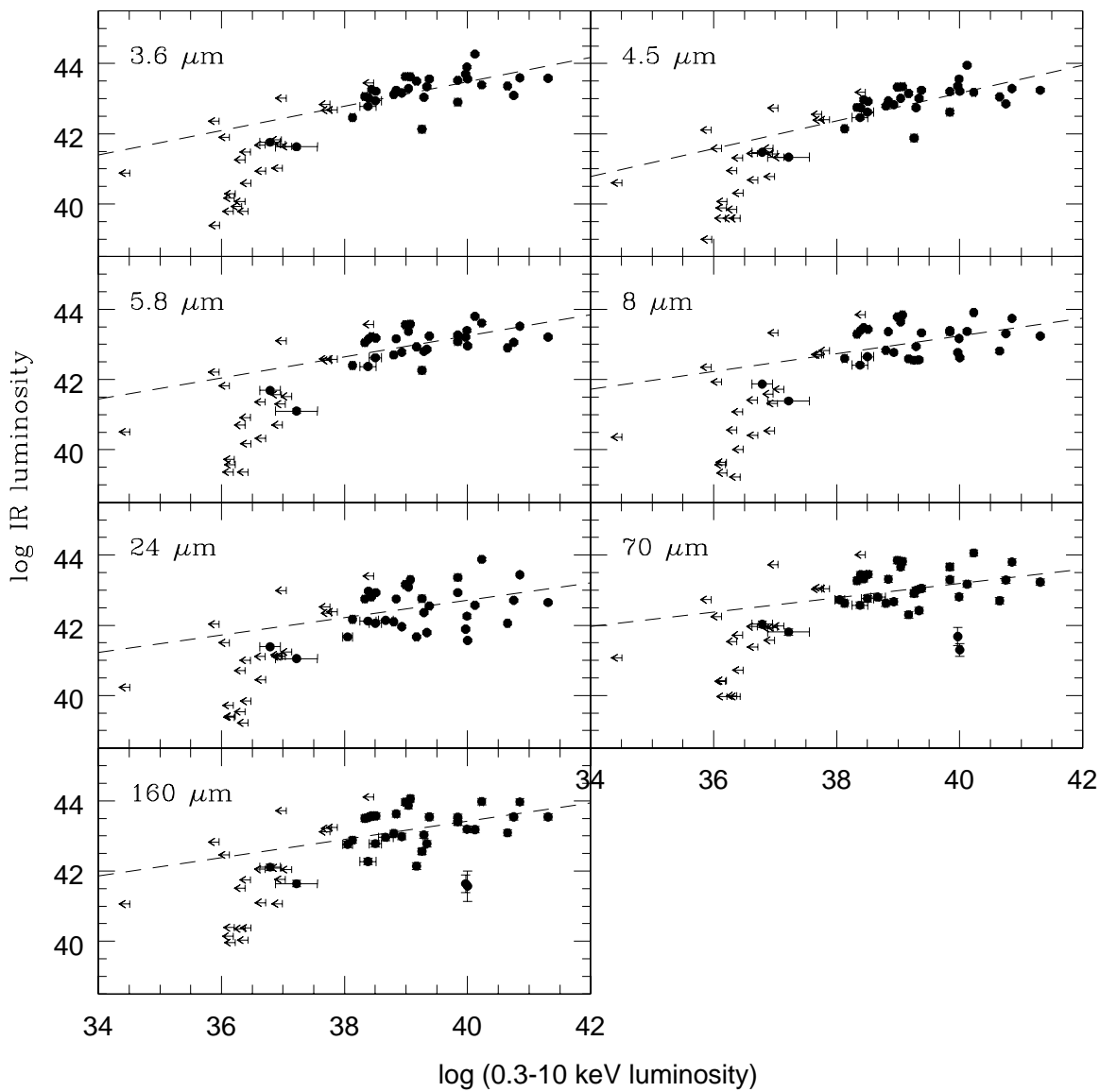}
\caption{$Spitzer$ IRAC and MIPS Luminosities of galaxies vs. Nuclear
(i.e. AGN) X-ray Luminosity. The dashed line represents the best fit
line to the data. Filled circles represent objects in which we have
detected the nuclear X-ray source; arrows represent upper limits for
those objects whose nuclei were not detected in X-rays.}
\label{fig:f9}
\end{center}
\end{figure}

As mentioned above, we obtained $Spitzer$ IRAC and MIPS flux densities
from \cite{Dale07} to see how the X-ray activity is related to other
galaxy properties. We converted these flux densities to observed
luminosities; since these objects are all at very low redshift, this
did not require a K-correction. Figure~\ref{fig:f9} shows the
$Spitzer$ luminosities of the galaxies plotted against the nuclear
(i.e. AGN) X-ray luminosities from this study. We observe correlations
between the nuclear X-ray luminosity and mid-infrared luminosity at
all wavelengths, but the correlation is the strongest at 3.6, 4.5 and
5.8 $\mu$m. The measured ratios and intercepts for all of the
correlations observed are given in Table \ref{Table:tbl5}. We also
compute the Spearman and Pearson's coefficients in each case to
determine the strength of the correlation. In both cases, a
coefficient of 1 represents a perfect linear correlation between the
two parameters, a coefficient of 0 represents no correlation, and
negative coefficients show a correlation in the opposite
direction. The coefficients are presented in Table
\ref{Table:tbl5}. We also calculated the probability that the the
observed correlation is not real, $P(r)$. The correlations between
nuclear X-ray luminosity and infrared luminosities at 3.6, 4.5, and
5.8 $\mu$m are significant at greater than 99.9\%; the correlation at
8 $\mu$m is significant to 99\%, and those at 24, 70, and 160 $\mu$m,
while much weaker than the other correlations, are still correlated at
$\sim$90\% certainty.  To ensure that the observed correlations are
not artifacts of sample selection or flux limits, we have also
included upper limits of $Chandra$ non-detections in
Figure~\ref{fig:f9}. It is clear that the observed correlations are
robust.

Note that our $Spitzer$ flux measurements measure the $global$
infrared fluxes-- not just the nuclear fluxes-- so comparing 3.6
$\mu$m and nuclear X-ray luminosity does not simply compare AGN
luminosities in two different bands unless the nuclear infrared flux
completely dominates the infrared flux of the entire galaxy. To be
sure that the nuclei of these galaxies were not dominating the global
flux in the infrared, we obtained nuclear surface brightness estimates
from \cite{Munozmateos09a}, which allowed us to estimate the total
flux within a 6\arcsec \ radius of the nucleus of each galaxy. We then
subtracted this flux from the global IR flux to see how much the total
was affected by removing the nuclear component, and in nearly all
cases, the removal of the nuclear IR flux did not have a significant
effect. This indicates that our global infrared fluxes do in fact
represent the flux of the galaxy rather than the AGN. Light at these
wavelengths is mostly contributed by stars; the observed correlation
between nuclear X-ray luminosity and IR luminosity is therefore a
manifestation of an observed relation between AGN luminosity and host
galaxy mass. The relationship between \mbh and stellar mass in the
host bulge has been well-established (e.g. \citealt{Kormendy95},
\citealt{Magorrian98}, \citealt{Bentz09a}), as has the relationship
between the size of the host bulge and the total stellar mass in the
galaxy (\citealt{Ferrarese02}, \citealt{Baes03}). Therefore it would
follow that \mbh correlates with the host galaxy stellar mass. Our
observed correlation between X-ray luminosity and host galaxy stellar
mass suggests that accretion rate ($\dot{M}$) also depends on the host
galaxy mass; this is a surprising new discovery.

\subsection{2MASS $J$, $H$, and $K$ band Measurements}
\begin{figure}
\begin{center}
\epsscale{1.0}
\plotone{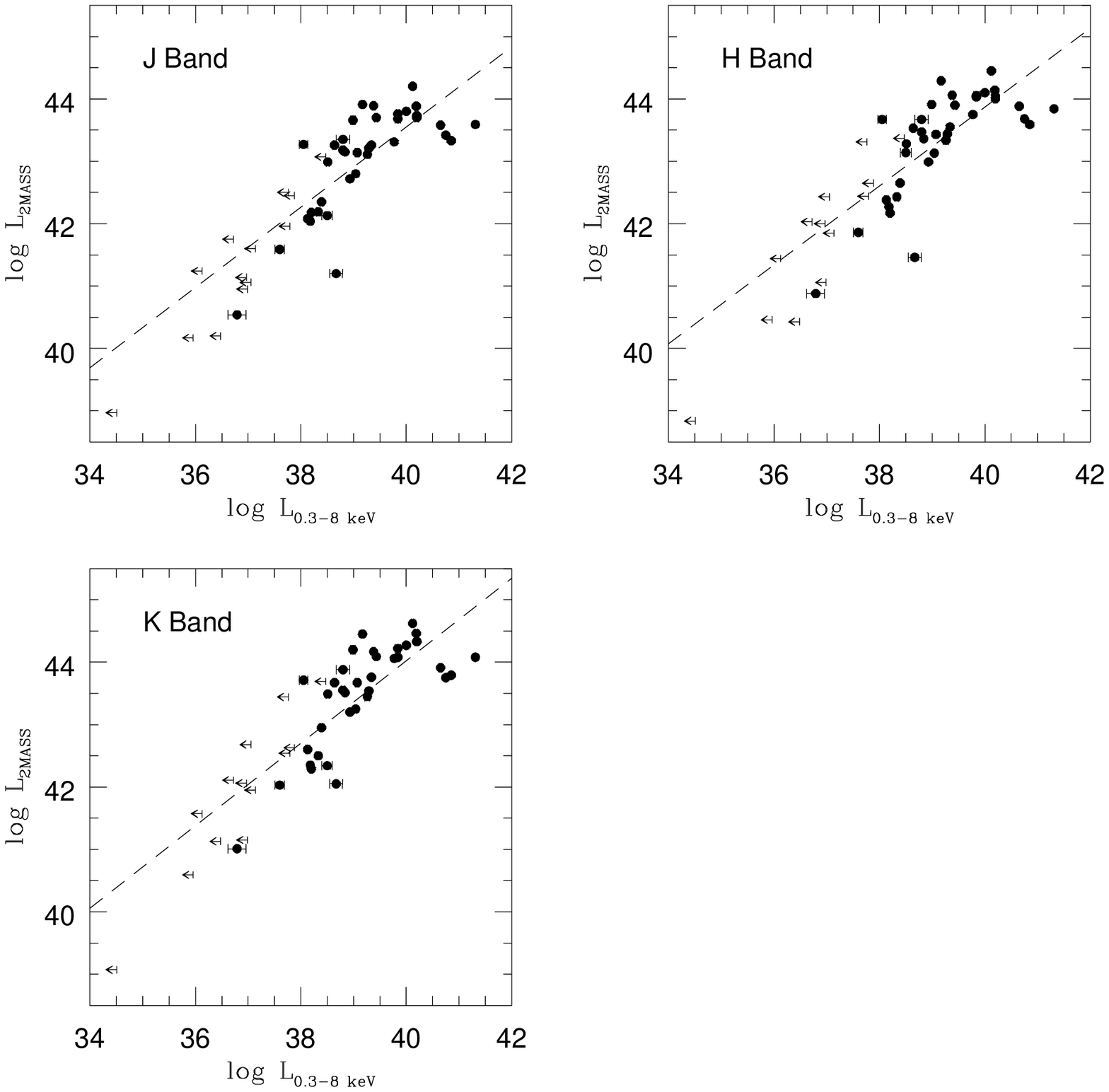}
\caption{ Nuclear 2MASS Luminosities vs. Nuclear X-ray Luminosity. The
dashed line represents the best fit line to the data. Filled circles
represent objects in which we have detected the nuclear X-ray source;
arrows represent upper limits for those objects whose nuclei were not
detected.}
\label{fig:f10}
\end{center}
\end{figure}

We looked for potential relations between nuclear 2MASS
luminosities and nuclear X-ray luminosities. As demonstrated in
Figure~\ref{fig:f10}, we observe trends with nuclear X-ray luminosity
that are well-fit by power laws; the best-fit parameters and
correlation coefficients are again given in Table
\ref{Table:tbl5}. The correlation coefficients indicate that the
correlations are significant in all three 2MASS bandpasses shown. Such
correlations should exist between X-ray and IR luminosities of AGNs;
observations of these correlations further support the identification
of nuclear X-ray sources as AGNs. The scatter around the correlations
and the non-unity slopes also suggest varying amounts of obscuration,
dust or intrinsic variations in the SED.

\begin{figure}
\begin{center}
\epsscale{1.0}
\plotone{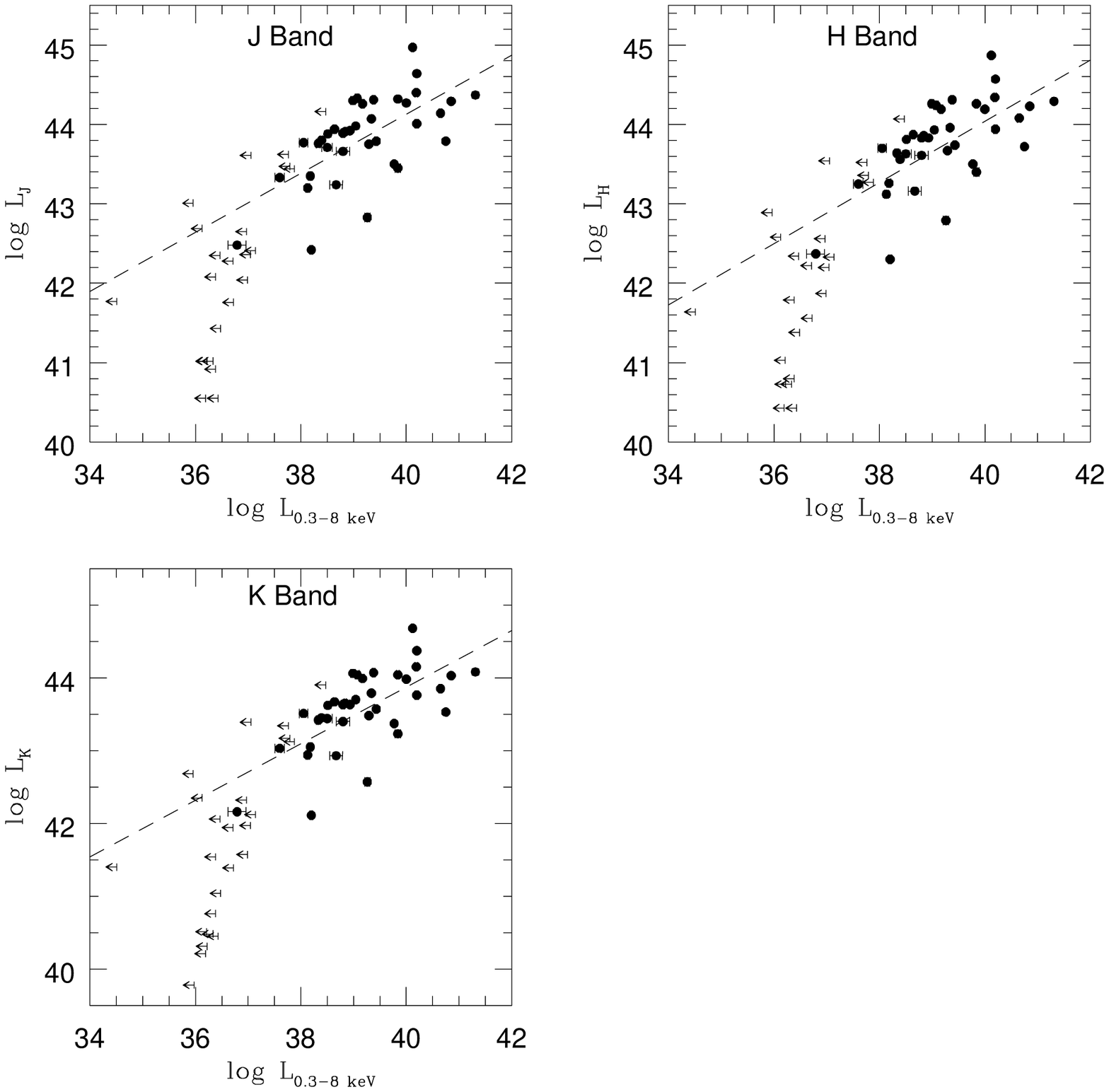}
\caption{Global 2MASS luminosities (from \citealt{Dale07}) vs. Nuclear
(i.e. AGN) X-ray Luminosity. The dashed line represents the best fit
line to the data. Filled circles represent objects in which we have
detected the nuclear X-ray source; arrows represent upper limits for
those objects whose nuclei were not detected in X-rays.}
\label{fig:f11}
\end{center}
\end{figure}

We also obtained global $J$, $H$, and $K$-band fluxes from
\cite{Dale07}. On a global scale, $K$ band magnitudes are good stellar
mass indicators, so we looked to see if there was a correlation
between nuclear X-ray luminosity and global $K$ band
luminosity. Figure \ref{fig:f11} shows this correlation; while
shallower than the $L_{K, \rm nuc}/L_{X}$ relation, it is still a
rather strong correlation. As with the $Spitzer$ data, we checked to
be sure that the $K$ band emission is not dominated by the nuclei,
and in most cases it did not. We removed the galaxies in which the
nuclear $K$ band flux was greater than 50\% of the total flux and
determined that these objects did not affect the measured correlation.

\begin{figure}
\begin{center}
\epsscale{0.5}
\plotone{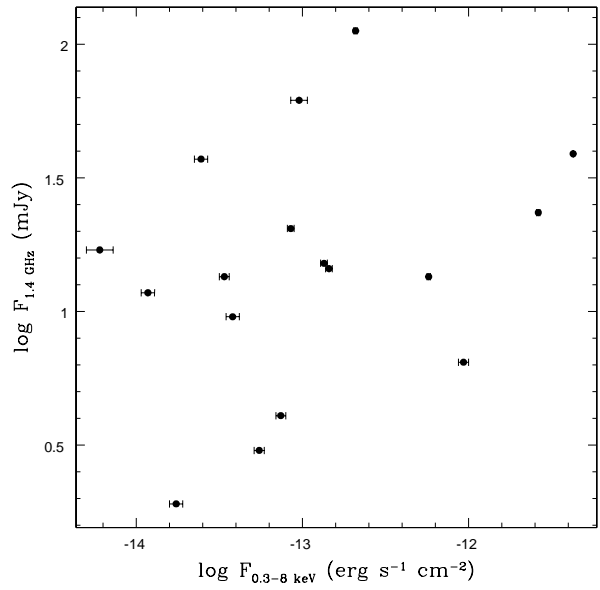}
\caption{Nuclear NVSS \& FIRST 1.4 GHz Integrated Flux densities
vs. Nuclear X-ray Flux}
\label{fig:f12}
\end{center}
\end{figure}

In addition, we examined the 1.4 GHz fluxes from the FIRST survey
(Figure~\ref{fig:f12}) and observe no correlation between the nuclear
X-ray flux and the integrated radio flux.

\subsection{Star Formation Rates}
\begin{figure} 
\begin{center}
\epsscale{1.0}
\plotone{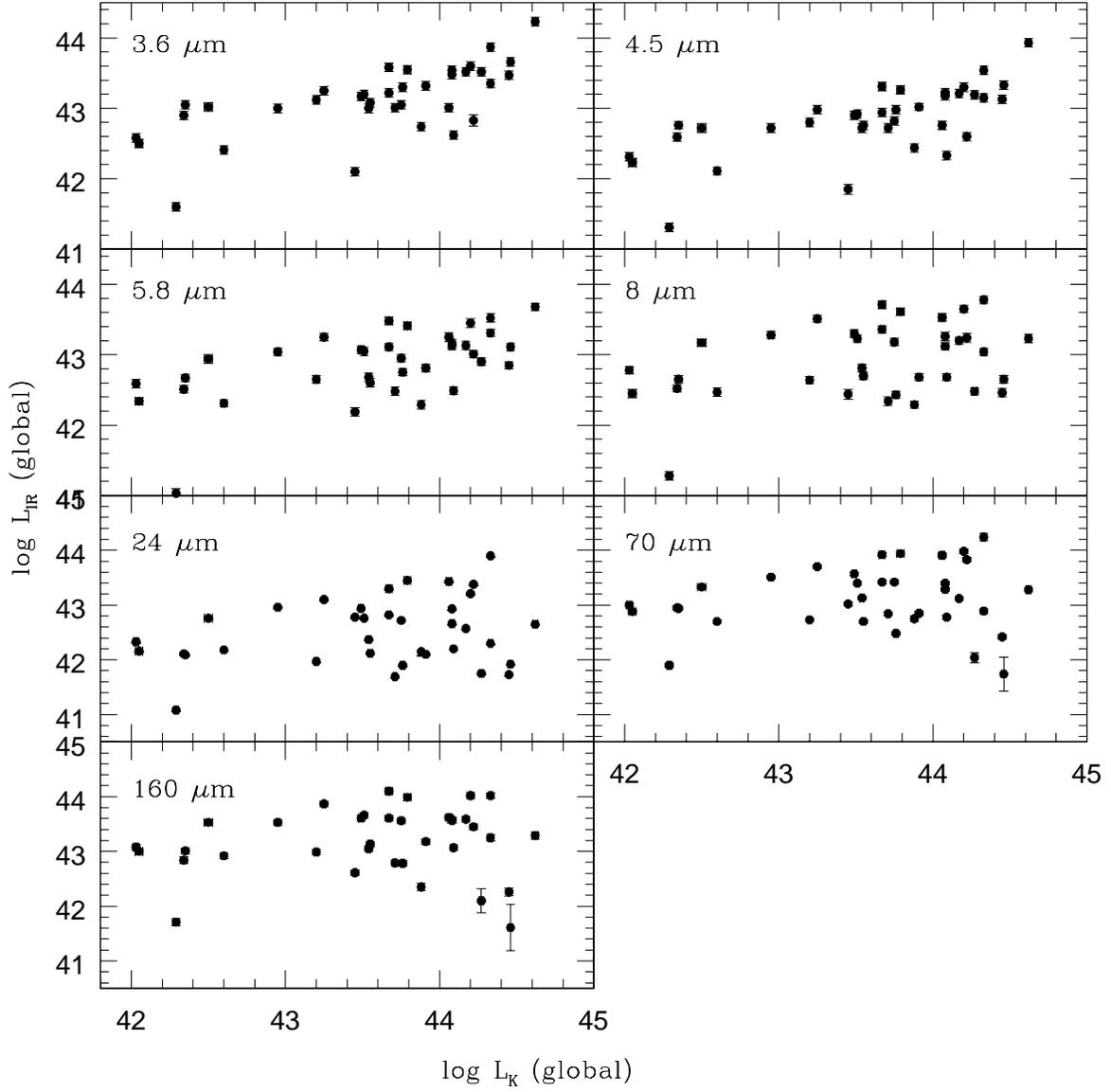}
\caption{Global $Spitzer$ IRAC and MIPS Luminosities vs. Global 2MASS
K-Band Luminosities (from \citealt{Dale07}).}
\label{fig:f13}
\end{center}
\end{figure}

Several studies in recent years have investigated a possible
connection between star formation and AGN luminosity in active
galaxies and quasars (e.g. \citealt{Netzer09b},
\citealt{Lutz08}). These correlations have been measured primarily
using star formation rate (SFR) indicators in the FIR, \Halpha \ and
\oii \ fluxes, or features that arise from polycyclic aromatic
hydrocarbons (PAHs) that are measured in the
mid-infrared. \cite{Netzer09} and \cite{Lutz08} both measure slopes of
around 0.8 when comparing the bolometric AGN luminosity (using either
$L_{5100}$ or [ \oi$/$\oii ] \ to estimate $L_{\rm bol}$) to 60 $\mu$m
luminosity. While we do not have the same observables, we can check
whether the galaxies in our sample show a correlation between AGN
X-ray luminosity and the 70 $\mu$m luminosity. We find a slope of
$\sim 0.2$ when comparing $L_{0.3-8 \rm keV}$ to $L_{70 \mu \rm
m}$. Since the AGNs are likely to be obscured, perhaps the X-ray
luminosity is not a perfect indicator of the total AGN luminosity; the
broad correlation between $L_{K, \rm nuc}$ and $L_{0.3-8 \rm keV}$
mentioned above, however, suggests that this is not a bad
assumption. A comparison between nuclear K-band luminosity $L_{K, \rm
nuc}$ and global $L_{70 \mu \rm m}$ results in a slope of 0.2 (see
Figure \ref{fig:f13}). Similarly shallow slopes are observed for
correlations between $L_{0.3-8 \rm keV}$ and other FIR measurements
(Table \ref{Table:tbl5}). Thus, regardless of whether we use X-ray,
K-band, or other IR luminosities as indicators of AGN activity, we do
not see a strong correlation of AGN activity with star formation
activity as seen at higher redshifts.

\begin{figure}
\begin{center}
\epsscale{0.5}
\plotone{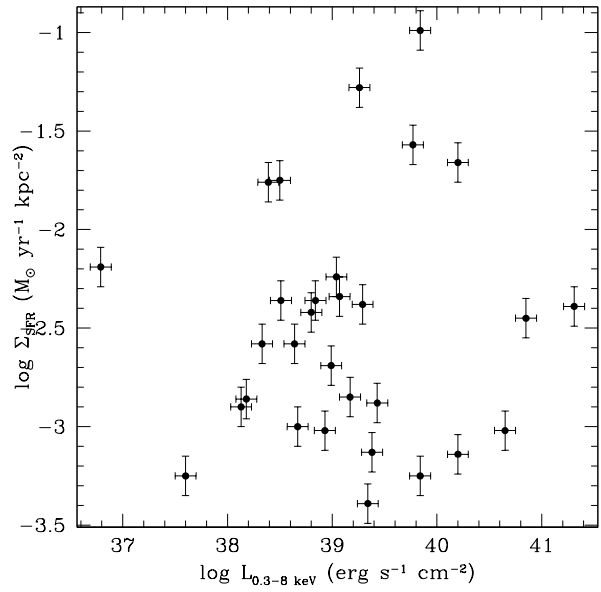}
\caption{Star formation rate per unit area from
\cite{Calzetti10}. These were measured using the $24\mu m$ luminosity
and H$\alpha$ luminosity as SFR indicators and are plotted against the nuclear X-ray
flux measured in this work.}
\label{fig:f14}
\end{center}
\end{figure}

We also examined the SFRs in our sample of nuclear X-ray detected
galaxies using SFRs computed by \cite{Calzetti10}. The SFRs were
calculated using the 24 $\mu$m luminosity combined with the H$\alpha$
luminosity as a composite star formaton rate indicator; see
\cite{Calzetti10} for details in these calculations.
Figure~\ref{fig:f14} shows the star formation rate per unit area in
our sample. We observe no significant correlation between nuclear
X-ray luminosity and SFR, which again is no surprise, as we do not
observe a strong correlation between $L_{0.3-8 \rm keV}$ and $L_{24
\mu \rm m}$ and therefore do not expect SFRs that are calculated using
$L_{24 \mu \rm m}$ to correlate either. Total SFR (rather than SFR per
unit area) was also examined and shows no significant correlation with
$L_{0.3-8 \rm keV}$.

\section {DISCUSSION}

In a sample of 62 SINGS galaxies with available $Chandra$ archival
data, we detect 37 nuclear X-ray sources. Eleven detections are in
objects whose nuclei have been identified as LINERs, five detections
are in identified \hii \ regions, twelve are in galaxies hosting
Seyfert nuclei, and two occurred in starburst galaxies. The other
seven detections were in galaxies which do not have a previous nuclear
classification. We argue that most of these 37 detections are likely
low-luminosity AGNs. We present nuclear X-ray fluxes, luminosities,
and hardness ratios for all of these sources. We look for connections
between infrared and X-ray luminosities and observe a correlation
between nuclear X-ray luminosity and various IR luminosities,
including IRAC 3.6 $\mu$m and 4.5 $\mu$m bands and 2MASS $J$, $H$, and
$K$ bands. Unlike the results for higher redshift AGNs, we find that
the AGN activity is {\it not} strongly correlated with the star
formation activity, though a mild trend is observed. We find instead a
strong correlation with the stellar mass of a galaxy. As discussed
above, it is well known that the mass of the nuclear BH is correlated
with the bulge mass or perhaps the total galaxy mass. Our observations
suggest that even the accretion rate depends on the galaxy mass. It is
possible that all of the galaxies are emitting at a similar fraction
of their Eddington luminosity; galaxies with higher mass BHs would
then be more luminous. New observations with $Herschel$
(\citealt{Shao10}) also show that the correlation between AGN activity
and star formation activity disappears at lower AGN luminosity even at
higher redshift. It thus appears that the AGN-SFR correlation is
not as generic a phenomenon as once thought.

\begin{figure}
\begin{center}
\epsscale{1.0}
\plotone{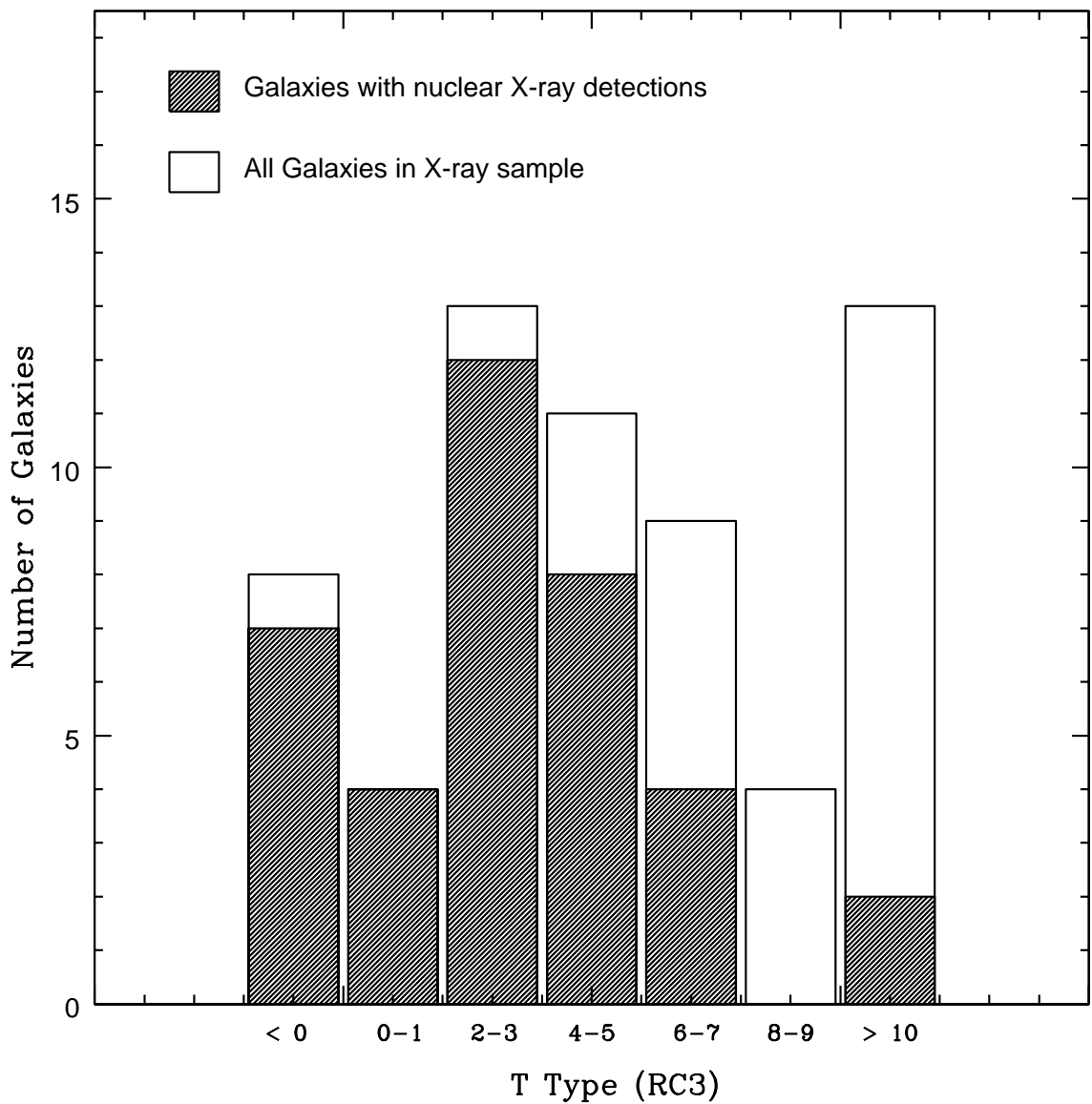}
\caption{Number of galaxies vs. Hubble ($T$) type. The shaded histogram
represents objects in which we have detected the nucleus; the unshaded
histogram represents all of the galaxies in the sample. Negative $T$
types represent elliptical and S0 galaxies; $T$ = 0 and 1 represent S0/a
and Sa galaxies; $T$ = 2 and 3 are Sab and Sb, $T$ = 4 and 5 are Sbc and Sc
galaxies, $T$ = 6 and 7 are Scd and Sd galaxies; $T$ = 8 and 9 are Sdm and
Sm galaxies, and $T$ $\geq$ 10 are irregulars and peculiars.}
\label{fig:f15}
\end{center}
\end{figure}

Figure~\ref{fig:f15} shows the distribution of Hubble types among the
SINGS galaxy sample in our study. We show the number of galaxies at
each Hubble type $T$ and compare this with the number of galaxies with
nuclear X-ray detections at each $T$ type. We observe nuclear X-ray
activity in nearly all of the galaxies towards the early end of the
Hubble sequence, and report no detections in Sdm and Sm galaxies and
only two nuclear X-ray detections in irregular or peculiar
galaxies. We may expect such a result if the bulge-less galaxies do
not harbor a nuclear BH. However, this is unlikely to be the case, as
BHs in bulge-less galaxies have been detected through optical
(\citealt{Peterson05}; \citealt{Shields08}), IR (\citealt{Satyapal07},
2009) and X-ray (\citealt{Ghosh08}) studies. Once again, it may
suggest lower accretion rates.

\begin{figure}
\begin{center}
\epsscale{1.0}
\plotone{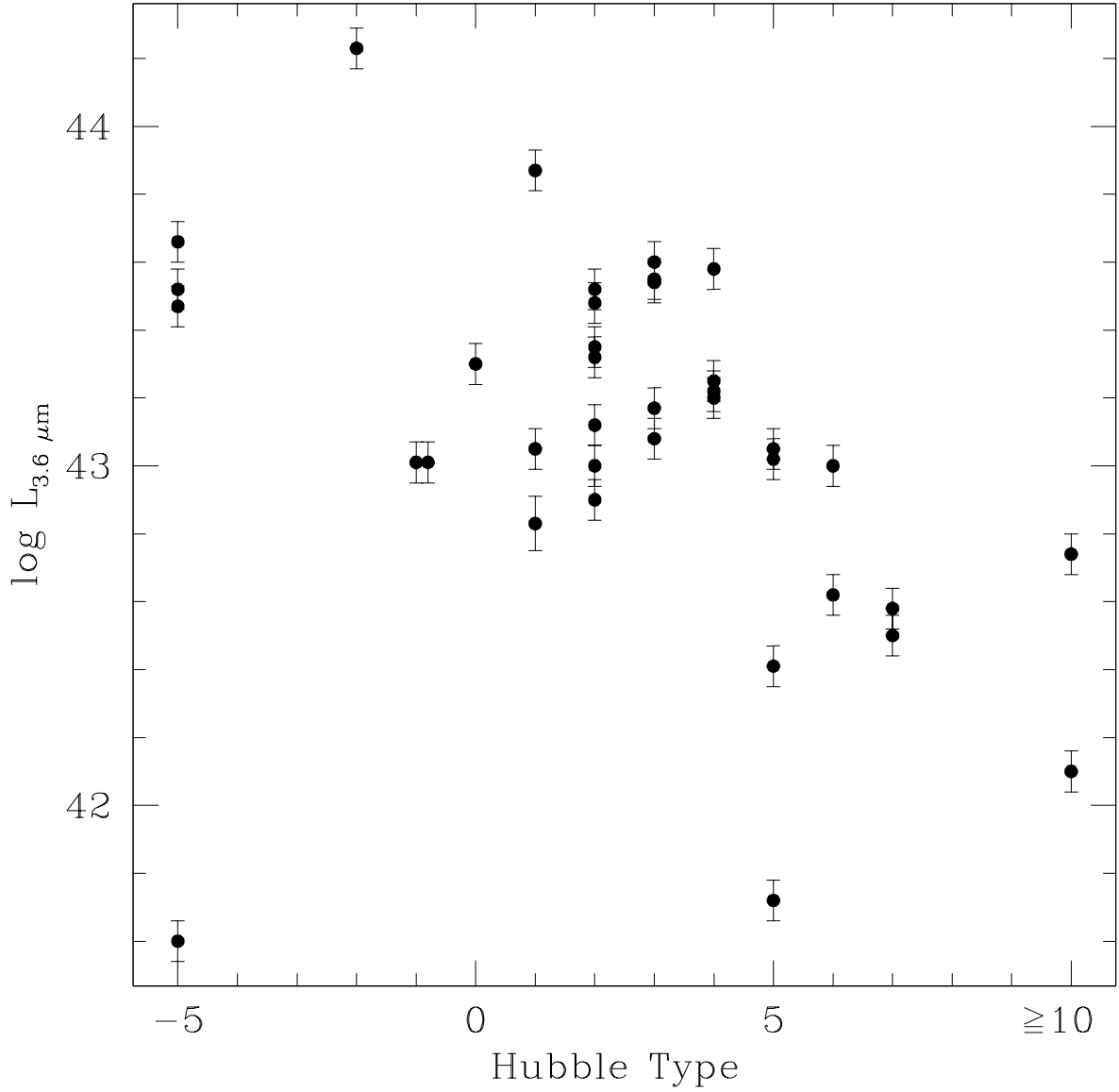}
\caption{Global $3.6 \mu m$ luminosity vs. Hubble Type. Corresponding
morphological types are listed in Figure \ref{fig:f15}.}
\label{fig:f16}
\end{center}
\end{figure}

We also show the 3.6\ $\mu$m luminosity of SINGS galaxies as a
function of Hubble Type in Figure~\ref{fig:f16}. As one would expect,
earlier-type galaxies show a higher 3.6 $\mu$m luminosity than
later-type galaxies, reflecting the differences in stellar mass
between early and late-type galaxies. Figure \ref{fig:f16} shows that
NGC\,855 is a significant outlier on the relation. We re-checked the
classification of the source and found that there is some disagreement
between sources over what type of galaxy this is. While it is listed
in the RC3 catalog as an elliptical (\citealt{Devaucouleurs91}), other
sources suggest that this galaxy is actually a dwarf elliptical
(e.g. \citealt{Roussel07}) or even a late-type spiral
(\citealt{Phillips96}). Our observed correlation supports the
reclassification of NGC\,855 as a dwarf elliptical.

\begin{figure}
\begin{center}
\epsscale{1.0}
\plotone{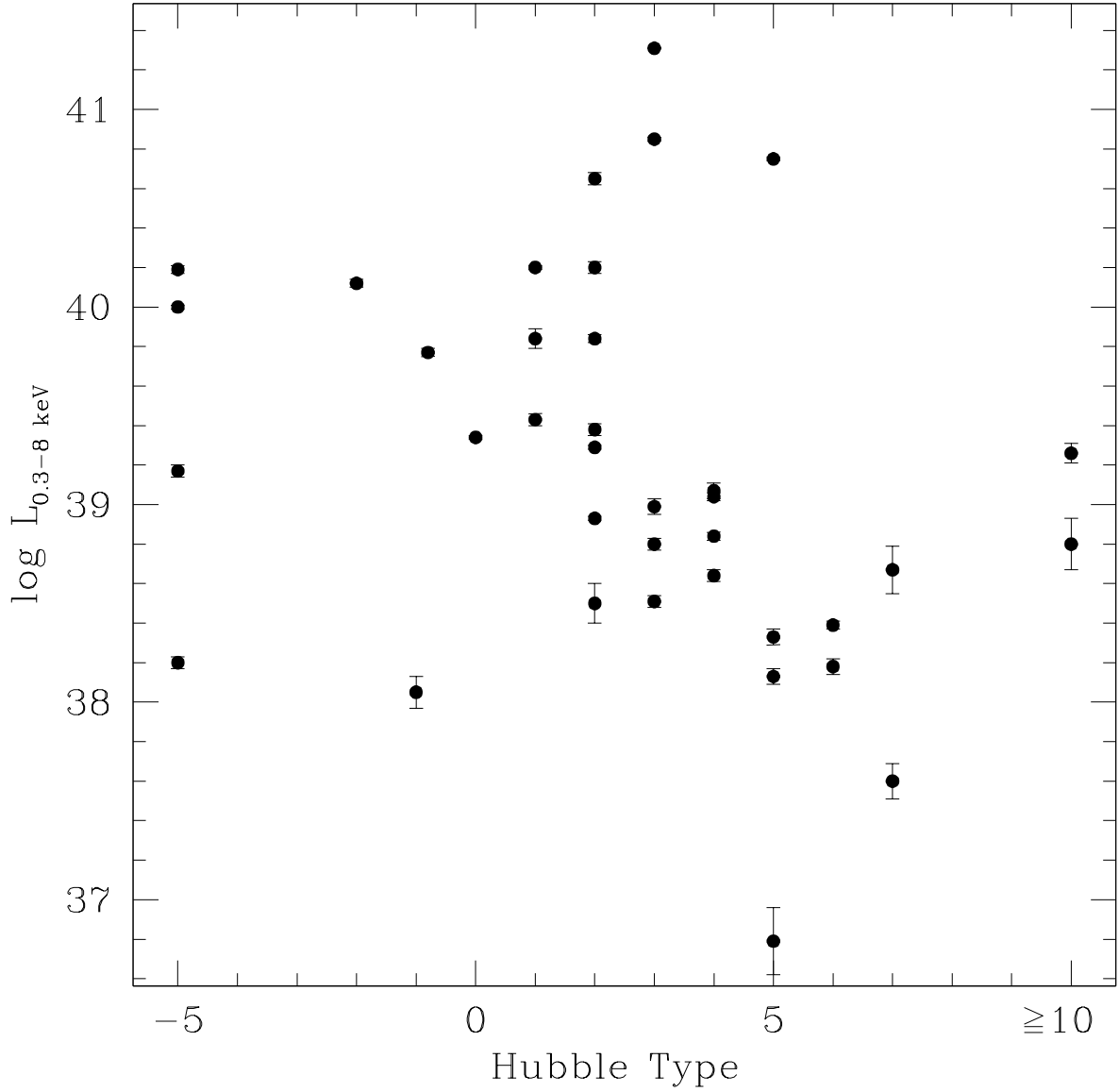}
\caption{Nuclear (AGN) X-ray Luminosity vs. Hubble Type. Corresponding
morphological types are listed in Figure \ref{fig:f15}.}
\label{fig:f17}
\end{center}
\end{figure}

Given the anticorrelation of $3.6 \mu$m luminosity with the Hubble
type and the correlation between $3.6 \mu$m luminosity with X-ray
luminosity, we expect to see an anticorrelation between the X-ray
luminosity and the Hubble type. However, as shown in
Figure~\ref{fig:f17}, there appears to be no correlation between these
measurements. This is likely because for every Hubble type there is a
wide range of $3.6 \mu$m luminosities and also a wide range of X-ray
luminosities.  We do notice that none of the galaxies later than $T$ = 5
have particularly high $L_{0.3-8 \rm keV}$.

\section {CONCLUSION}

We find that about 60\% of SINGS galaxies host nuclear X-ray sources
which are likely to be AGNs. This fraction is much larger than that
found through optical studies (e.g. \citealt{Ho97}) and shows the
efficacy of X-ray observations to find hidden AGNs in normal
galaxies. We find that for our sample of galaxies AGN activity is
correlated with the stellar mass of the galaxy. This is a surprising
new result and suggests that together with the mass of the SMBH,
accretion rate also depends on galaxy mass.  Unlike the merger driven
black hole growth observed at high redshift for high-luminosity AGN,
there appears to be an alternative mode of black hole growth at the
present epoch in late type galaxies.  It has been suggested in
literature that the total mass of a galaxy, not just the bulge mass,
is the primary driver of the mass of the SMBH (\citealt{Ferrarese02},
\citealt{Baes03}); our results support such a scenario.

\acknowledgments We are grateful to the SINGS team for the
multiwavelength survey of nearby galaxies. This publication makes use
of data products from the Two Micron All Sky Survey, which is a joint
project of the University of Massachusetts and the Infrared Processing
and Analysis Center/California Institute of Technology, funded by the
National Aeronautics and Space Administration and the National Science
Foundation. This study also made use of the VizieR database of
astronomical catalogues (\citealt{Ochsenbein00}) as well as the
NASA/IPAC Extragalactic Database (NED) which is operated by the Jet
Propulsion Laboratory, California Institute of Technology, under
contract with the National Aeronautics and Space Administration. This
research has also made use of data obtained from the Chandra Data
Archive and the Chandra Source Catalog, and software provided by the
Chandra X-ray Center (CXC) in the application packages CIAO, ChIPS,
and Sherpa.

This work is supported in part by the National Aeronautics and Space
Administration through Chandra award number GO7-8111X issued by the
Chandra X-ray Observatory Center, which is operated by the Smithsonian
Astrophysical Observatory for and on behalf of the National
Aeronautics and Space Administration under contract NAS8-03060.

\bibliographystyle{apjsym} 

\clearpage

%

%
%
%
%
%
%

\clearpage

\begin{deluxetable}{lccccccccr} 
\tabletypesize{\scriptsize}
\setlength{\tabcolsep}{0.04in}
\tablecaption{SINGS Galaxy Properties and References} 
\tablewidth{0pt} 
\tablehead{ 
\colhead{} & 
\colhead{RA} & 
\colhead{DEC} & 
\colhead{Morphological} & 
\colhead{T } & 
\colhead{Nuclear} & 
\colhead{Dist} & 
\colhead{} & 
\colhead{} &  \\ 
\colhead{Galaxy} & 
\colhead{(J2000)} & 
\colhead{(J2000)} & 
\colhead{Type (RC3)} & 
\colhead{(RC3)} & 
\colhead{Type\tablenotemark{a}} & 
\colhead{(Mpc)\tablenotemark{b}} & 
\colhead{R mag\tablenotemark{b}} & 
\colhead{Publications\tablenotemark{c}} \\ 
} 
\startdata
NGC\,0024	&	00 09 56.5	&	¡Ý24 57 47	&	SAc	&	5	&	\nodata	&	8.2	&	¡Ý18.4	&	1	\\
NGC\,0337	&	00 59 50.1	&	¡Ý07 34 41	&	SBd	&	7	&	\nodata	&	25	&	¡Ý20.3	&	\nodata	\\
NGC\,0584	&	01 31 20.7	&	¡Ý06 52 05	&	E4 	&	-5	&	\nodata	&	28	&	¡Ý22.2	&	\nodata	\\
NGC\,0628	&	01 36 41.8	&	+15 47 00	&	SAc	&	5	&	\nodata	&	11	&	¡Ý20.9	&	2,3, 4	\\
NGC\,0855	&	02 14 03.6	&	+27 52 38	&	E  	&	-5	&	\nodata	&	9.6	&	¡Ý17.7	&	2	\\
NGC\,0925	&	02 27 16.9	&	+33 34 45	&	SABd	&	7	&	H	&	10.1	&	¡Ý20.6	&	2,5	\\
NGC\,1097	&	02 46 19.0	&	¡Ý30 16 30	&	SBb	&	3	&	L	&	16.9	&	¡Ý22.4	&	6	\\
NGC\,1266	&	03 16 00.7	&	¡Ý02 25 38	&	SB0	&	-2	&	Sy	&	31.3	&	¡Ý21.1	&	\nodata	\\
NGC\,1291	&	03 17 18.6	&	¡Ý41 06 29	&	SBa	&	0	&	\nodata	&	9.7	&	¡Ý22.0	&	2,4, 7	\\
NGC\,1316	&	03 22 41.7	&	¡Ý37 12 30	&	SAB0	&	-2	&	L	&	26.3	&	¡Ý23.5	&	8, 9	\\
NGC\,1377	&	03 36 39.1	&	¡Ý20 54 08	&	S0 	&	-2	&	H	&	24.4	&	¡Ý19.6	&	\nodata	\\
NGC\,1404	&	03 38 51.9	&	¡Ý35 35 40	&	E1 	&	-5	&	\nodata	&	25.1	&	¡Ý22.9	&	1	\\
NGC\,1482	&	03 54 38.9	&	¡Ý20 30 09	&	SA0	&	-0.8	&	\nodata	&	22	&	¡Ý20.5	&	10	\\
NGC\,1512	&	04 03 54.3	&	¡Ý43 20 56	&	SBab	&	1	&	SB	&	10.4	&	¡Ý19.9	&	\nodata	\\
NGC\,1566	&	04 20 00.4	&	¡Ý54 56 16	&	SABbc	&	4	&	Sy1.5	&	18	&	¡Ý21.9	&	\nodata	\\
NGC\,1705	&	04 54 13.5	&	¡Ý53 21 40	&	Am	&	-3	&	SB	&	5.8	&	¡Ý16.7	&	2	\\
NGC\,2403	&	07 36 51.4	&	+65 36 09	&	SABcd	&	6	&	H	&	3.5	&	¡Ý19.7	&	2,5, 11	\\
Ho II	&	08 19 05.0	&	+70 43 12	&	Im	&	10	&	\nodata	&	3.5	&	¡Ý17.1	&	\nodata	\\
M81 DwA	&	08 23 55.1	&	+71 01 56	&	Im	&	10	&	\nodata	&	3.5	&	¡Ä	&	1	\\
DDO 053	&	08 34 07.2	&	+66 10 54	&	Im	&	10	&	\nodata	&	3.5	&	¡Ý13.6	&	1	\\
NGC\,2798	&	09 17 22.9	&	+41 59 59	&	SBa	&	1	&	SB	&	24.7	&	¡Ý19.6	&	1	\\
NGC\,2841	&	09 22 02.6	&	+50 58 35	&	SAb	&	3	&	Sy3	&	9.8	&	¡Ý20.7	&	12, 13	\\
NGC\,2915	&	09 26 11.5	&	¡Ý76 37 35	&	I0	&	90	&	SB	&	2.7	&	¡Ý15.1	&	1	\\
Ho I	&	09 40 32.3	&	+71 10 56	&	IABm	&	10	&	\nodata	&	3.5	&	¡Ý13.2	&	1	\\
NGC\,2976	&	09 47 15.4	&	+67 54 59	&	SAc	&	5	&	H	&	3.5	&	¡Ý17.6	&	1	\\
NGC\,3049	&	09 54 49.5	&	+09 16 16	&	SBab	&	2	&	SB	&	19.6	&	¡Ý18.7	&	\nodata	\\
NGC\,3031 (M81)	&	09 55 33.2	&	+69 03 55	&	SAab	&	2	&	L	&	3.5	&	¡Ý21.2	&	2,13, 14	\\
NGC\,3034 (M82)	&	09 55 52.7	&	+69 40 46	&	IO	&	90	&	SB	&	3.5	&	¡Ý17.9	&	2	\\
Ho IX	&	09 57 32.0	&	+69 02 45	&	Im	&	10	&	\nodata	&	3.5	&	¡Ý13.6	&	2	\\
M81 DwB	&	10 05 30.6	&	+70 21 52	&	Im	&	10	&	\nodata	&	3.5	&	¡Ý12.5	&	1	\\
NGC\,3190	&	10 18 05.6	&	+21 49 55	&	SAap	&	1	&	L	&	17.4	&	¡Ý20.7	&	15, 16	\\
NGC\,3184	&	10 18 17.0	&	+41 25 28	&	SABcd	&	6	&	H	&	8.6	&	¡Ý19.0	&	2,5, 17	\\
NGC\,3198	&	10 19 54.9	&	+45 32 59	&	SBc	&	5	&	\nodata	&	9.8	&	¡Ý20.2	&	1	\\
IC 2574	&	10 28 23.5	&	+68 24 44	&	SABm	&	9	&	\nodata	&	3.5	&	¡Ý17.7	&	2	\\
NGC\,3265	&	10 31 06.8	&	+28 47 48	&	E	&	-5	&	\nodata	&	20	&	¡Ý17.7	&	\nodata	\\
Mrk 33	&	10 32 31.9	&	+54 24 04	&	Im	&	10	&	SB	&	21.7	&	¡Ý18.4	&	1	\\
NGC\,3351	&	10 43 57.7	&	+11 42 14	&	SBb	&	3	&	SB	&	9.3	&	¡Ý20.4	&	2,18	\\
NGC\,3521	&	11 05 48.6	&	¡Ý00 02 09	&	SABbc	&	4	&	L	&	9	&	¡Ý21.0	&	2	\\
NGC\,3621	&	11 18 16.5	&	-32 48 51	&	Sad	&	7	&	Sy2	&	6.2	&	¡Ý19.4	&	19	\\
NGC\,3627	&	11 20 15.0	&	+12 59 30	&	SABb	&	3	&	Sy3	&	8.9	&	¡Ý20.8	&	2,13	\\
NGC\,3773	&	11 38 12.9	&	+12 06 43	&	SA0	&	-2	&	\nodata	&	12.9	&	¡Ý17.5	&	\nodata	\\
NGC\,3938	&	11 52 49.4	&	+44 07 15	&	SAc	&	5	&	\nodata	&	12.2	&	¡Ý20.1	&	1	\\
NGC\,4125	&	12 08 06.0	&	+65 10 27	&	E6p	&	-5	&	\nodata	&	21.4	&	¡Ý21.6	&	11,20, 21	\\
NGC\,4236	&	12 16 42.1	&	+69 27 45	&	SBdm	&	8	&	\nodata	&	3.5	&	¡Ý18.1	&	1	\\
NGC\,4254	&	12 18 49.6	&	+14 24 59	&	SAc	&	5	&	\nodata	&	20	&	¡Ý21.6	&	4	\\
NGC\,4321	&	12 22 54.9	&	+15 49 21	&	SABbc	&	4	&	L	&	20	&	¡Ý22.1	&	2,12, 13	\\
NGC\,4450	&	12 28 29.6	&	+17 05 06	&	SAab	&	2	&	L	&	20	&	¡Ý21.4	&	3	\\
NGC\,4536	&	12 34 27.0	&	+02 11 17	&	SABbc	&	4	&	H	&	25	&	¡Ý20.8	&	\nodata	\\
NGC\,4552	&	12 35 39.8	&	+12 33 23	&	E0	&	-5	&	Sy2	&	20	&	¡Ý20.8	&	20, 22, 23, 3, 12,24,25	\\
NGC\,4559	&	12 35 57.7	&	+27 57 35	&	SABcd	&	6	&	H	&	11.6	&	¡Ý21.0	&	2,5	\\
NGC\,4569	&	12 36 49.8	&	+13 09 46	&	SABab	&	2	&	Sy	&	20	&	¡Ý22.0	&	13, 21	\\
NGC\,4579	&	12 37 43.5	&	+11 49 05	&	SABb	&	3	&	L	&	20	&	¡Ý21.8	&	13, 21, 12,26	\\
NGC\,4594	&	12 39 59.4	&	¡Ý11 37 23	&	SAa	&	1	&	Sy3	&	13.7	&	¡Ý21.5	&	2, 12, 13	\\
NGC\,4625	&	12 41 52.7	&	+41 16 26	&	SABmp	&	9	&	Sy	&	9.5	&	¡Ý17.5	&	2,5	\\
NGC\,4631	&	12 42 08.0	&	+32 32 29	&	SBd	&	7	&	\nodata	&	9	&	¡Ý20.6	&	2,5	\\
NGC\,4725	&	12 50 26.6	&	+25 30 03	&	SABab	&	2	&	Sy2	&	17.1	&	¡Ý22.0	&	2,13	\\
NGC\,4736	&	12 50 53.0	&	+41 07 14	&	SAab	&	2	&	Sy	&	5.3	&	¡Ý19.9	&	2,12,24,26	\\
DDO 154	&	12 54 05.2	&	+27 08 59	&	IBm	&	10	&	\nodata	&	5.4	&	¡Ý15.1	&	1	\\
NGC\,4826	&	12 56 43.7	&	+21 40 58	&	SAab	&	2	&	Sy2	&	5.6	&	¡Ý20.3	&	2,13	\\
DDO 165	&	13 06 24.8	&	+67 42 25	&	Im	&	10	&	\nodata	&	3.5	&	¡Ý15.3	&	1	\\
NGC\,5033	&	13 13 27.5	&	+36 35 38	&	SAc	&	5	&	Sy2	&	13.3	&	¡Ý20.9	&	13,27	\\
NGC\,5055	&	13 15 49.3	&	+42 01 45	&	SAbc	&	4	&	H/L	&	8.2	&	¡Ý19.0	&	20,2,12,28	\\
NGC\,5194 (M51)	&	13 29 52.7	&	+47 11 43	&	SABbc	&	4	&	Sy2	&	8.2	&	¡Ý21.4	&	2,12,29,21	\\
NGC\,5195	&	13 29 59.6	&	+47 15 58	&	SB0p	&	90	&	L	&	8.2	&	¡Ý20.0	&	30,13,21,2,31	\\
NGC\,5398	&	14 01 21.5	&	¡Ý33 03 50	&	SBdm	&	8.1	&	H	&	15	&	¡Ý18.9	&	\nodata	\\
NGC\,5408	&	14 03 20.9	&	¡Ý41 22 40	&	IBm	&	90	&	\nodata	&	4.5	&	¡Ý16.1	&	1	\\
NGC\,5474	&	14 05 01.6	&	+53 39 44	&	SAcd	&	6	&	H	&	6.9	&	¡Ý18.4	&	2, 5	\\
NGC\,5713	&	14 40 11.5	&	¡Ý00 17 20	&	SABbcp	&	4	&	\nodata	&	26.6	&	¡Ý20.9	&	\nodata	\\
NGC\,5866 (M102)	&	15 06 29.5	&	+55 45 48	&	SA0	&	-1	&	Sy	&	12.5	&	¡Ý19.9	&	12, 20, 24, 32, 27	\\
IC 4710	&	18 28 38.0	&	¡Ý66 58 56	&	SBm	&	9	&	SB	&	8.5	&	¡Ý18.3	&	\nodata	\\
NGC\,6822	&	19 44 56.6	&	¡Ý14 47 21	&	IBm	&	10	&	\nodata	&	0.6	&	¡Ý13.8	&	2	\\
NGC\,6946	&	20 34 52.3	&	+60 09 14	&	SABcd	&	6	&	H	&	5.5	&	¡Ý21.3	&	5, 2, 33	\\
NGC\,7331	&	22 37 04.1	&	+34 24 56	&	SAb	&	3	&	L	&	15.7	&	¡Ý21.8	&	2, 12, 20, 21	\\
NGC\,7552	&	23 16 10.8	&	¡Ý42 35 05	&	SAc	&	2	&	H	&	22.3	&	¡Ý21.7	&	4	\\
NGC\,7793	&	23 57 49.8	&	¡Ý32 35 28	&	SAd	&	7	&	H	&	3.2	&	¡Ý18.2	&	5	\\

\enddata
\tablenotetext{a}{Nuclear identifications: S=Seyfert, SB= starburst, H= nuclear \hii region, L= Liner nucleus. Most values are from the RC3 catalog 
and have been supplemented by values from \cite{Veroncetty10}.}
\tablenotetext{b}{From \cite{Kennicutt03}}
\tablenotetext{c}{Publications containing analyses of galactic nuclei using $Chandra$ data. References:
(1) This paper; 
(2) \citealt{Zhang09}; 
(3) \citealt{Ho09}; 
(4) \citealt{Ghosh09phd};
(5) \citealt{Desroches09}; 	   
(6) \citealt{Nemmen06}; 	   
(7) \citealt{Irwin02};		   
(8) \citealt{Kim03};		   
(9) \citealt{Rinn05};		   
(10 \citealt{Strickland04};	   
(11)\citealt{Schlegel03};	   
(12) \citealt{Gonzalez09}	   
(13) \citealt{Ho01};		   
(14) \citealt{Swartz03};	   
(15) \citealt{Hughes07};	   
(16) \citealt{Rasmussen08};	   
(17) \citealt{Ghosh08};		   
(18) \citealt{Swartz06};	   
(19) \citealt{Gliozzi09};	   
(20) \citealt{Flohic06};	   
(21) \citealt{Satyapal04};	   
(22) \citealt{Gallo08};		    	   
(23) \citealt{Xu05};		    
(24) \citealt{Satyapal05};	    
(25) \citealt{Filho04};		    
(26) \citealt{Eracleous02};	    
(27) \citealt{Terashima03};	    
(28) \citealt{Luo07};		    
(29) \citealt{Terashima01};	    
(30) \citealt{Georgantopoulos02};   
(31) \citealt{Terashima04};	    
(32) \citealt{Li09};		    
(33) \citealt{Holt03}.}             
                
\label{Table:tbl1}
\end{deluxetable} 

\begin{deluxetable}{lclccccccccccr} 
\tablewidth{0pt} 
\rotate
\tabletypesize{\scriptsize}
\setlength{\tabcolsep}{0.03in}
\tablecaption{Source Information for Nuclear Detections} 
\tablehead{ 
\colhead{} & 
\colhead{} & 
\colhead{Observation} &
\colhead{Exposure} &
\colhead{} & 
\colhead{Source} & 
\colhead{Source} & 
\colhead{Bkg/Src} & 
\colhead{Broad} & 
\colhead{Broad} & 
\colhead{Net} & 
\colhead{Net} & 
\colhead{Net} &
\colhead{Hardness} \\ 
\colhead{Object} & 
\colhead{OBSID} & 
\colhead{Date} &
\colhead{Time (ks)} & 
\colhead{CCD} & 
\colhead{RA} & 
\colhead{DEC} & 
\colhead{Ratio\tablenotemark{a}} & 
\colhead{Source} & 
\colhead{Bkg} & 
\colhead{Broad} & 
\colhead{Hard} & 
\colhead{Soft} &
\colhead{Ratio} \\ 
} 
\startdata
NGC\,0024	&	9547	&	13. October 2008	&	43.22	&	7	&	\nodata	&	\nodata	&	19.60	&	\nodata	&	42	&	$<$	7	&	$<$	6	&	$<$	4	&	\nodata	\\
NGC\,0628	&	2057	&	19. June 2001	&	43.2	&	7	&	01:36:41.74	&	15:47:01.19	&	19.73	&	97	&	42	&		95	&		22	&		73	&	-0.529	\\
NGC\,0855	&	9550	&	3. October 2008	&	58.73	&	7	&	02:14:03.50	&	27:52:38.40	&	18.92	&	108	&	98	&		103	&		33	&		69	&	-0.349	\\
NGC\,0925	&	7104	&	23. November 2005	&	2.24	&	7	&	02:27:16.87	&	33:34:45.30	&	20.91	&	13	&	4	&		13	&		0	&		13	&	\nodata	\\
NGC\,1097\tablenotemark{*}	&	2339	&	28. January 2001	&	4	&	7	&	02:46:18.97	&	-30:16:28.98	&	20.21	&	1263	&	382	&		1244	&		209	&		1035	&	-0.663	\\
NGC\,1291\tablenotemark{*}	&	795	&	18. July 2001	&	37.27	&	7	&	03:17:18.59	&	-41:06:28.72	&	17.40	&	1102	&	430	&		1077	&		421	&		656	&	-0.218	\\
NGC\,1316	&	2022	&	17. April 2001	&	25.85	&	7	&	03:22:41.69	&	-37:12:28.71	&	16.81	&	702	&	1483	&		614	&		57	&		557	&	-0.814	\\
NGC\,1404	&	4174	&	28. May 2003	&	43.46	&	3	&	03:38:51.93	&	-35:35:39.06	&	19.79	&	938	&	2782	&		797	&		38	&		759	&	-0.904	\\
NGC\,1482	&	2932	&	5. February 2002	&	24.72	&	7	&	03:54:39.32	&	-20:30:09.53	&	19.66	&	417	&	899	&		372	&		172	&		200	&	-0.077	\\
NGC\,1705	&	3930	&	12. September 2003	&	45.31	&	7	&	\nodata	&	\nodata	&	20.94	&	\nodata	&	76	&	$<$	9	&	$<$	4	&	$<$	8	&	\nodata	\\
NGC\,2403	&	4628	&	23. August 2004	&	46.21	&	7	&	\nodata	&	\nodata	&	20.94	&	\nodata	&	42	&	$<$	4	&	$<$	2	&	$<$	3	&	\nodata	\\
M81 DwA	&	9535	&	24. August 2008	&	25.92	&	7	&	\nodata	&	\nodata	&	20.94	&	\nodata	&	14	&	$<$	2	&	$<$	2	&	$<$	2	&	\nodata	\\
DDO 053	&	9538	&	3. February 2008	&	19.02	&	7	&	\nodata	&	\nodata	&	20.94	&	\nodata	&	21	&	$<$	3	&	$<$	2	&	$<$	2	&	\nodata	\\
NGC\,2798	&	10567	&	24. January 2009	&	5.11	&	7	&	09:17:22.74	&	42:00:00.12	&	21.00	&	76	&	64	&		73	&		7	&		66	&	-0.82	\\
NGC\,2841	&	6096	&	18. December 2004	&	26.95	&	7	&	09:22:02.68	&	50:58:35.78	&	19.75	&	230	&	193	&		220	&		36	&		184	&	-0.672	\\
NGC\,2915	&	9534	&	10. April 2008	&	15.36	&	7	&	\nodata	&	\nodata	&	20.94	&	\nodata	&	17	&	$<$	8	&	$<$	5	&	$<$	6	&	\nodata	\\
Ho I	&	9539	&	22. June 2008	&	25.92	&	7	&	\nodata	&	\nodata	&	20.94	&	\nodata	&	21	&	$<$	4	&	$<$	2	&	$<$	4	&	\nodata	\\
NGC\,2976	&	9542	&	24. March 2008	&	9.82	&	7	&	09:47:15.35	&	67:55:00.10	&	21.00	&	7	&	17	&		6	&		5	&		1	&	\nodata	\\
NGC\,3031\tablenotemark{*}	&	735	&	7. May 2000	&	49.07	&	7	&	09:55:33.37	&	69:03:53.49	&	20.94	&	4349	&	4421	&		4099	&		1413	&		2686	&	-0.311	\\
NGC\,3034 \tablenotemark{*}	&	8190	&	2. June 2007	&	48.99	&	7	&	\nodata	&	\nodata	&	14.20	&	\nodata	&	13967	&	$<$	138	&	$<$	71	&	$<$	118	&	\nodata	\\
Ho IX	&	9540	&	24. August 2008	&	25.72	&	7	&	\nodata	&	\nodata	&	20.94	&	\nodata	&	32	&	$<$	6	&	$<$	3	&	$<$	5	&	\nodata	\\
M81 DwB	&	9536	&	7. July 2008	&	24.93	&	7	&	\nodata	&	\nodata	&	20.60	&	\nodata	&	24	&	$<$	7	&	$<$	4	&	$<$	6	&	\nodata	\\
NGC\,3190\tablenotemark{*}	&	2760	&	14. March 2002	&	19.81	&	7	&	10:18:05.63	&	21:49:56.39	&	21.00	&	225	&	98	&		220	&		75	&		145	&	-0.317	\\
NGC\,3184	&	804	&	8. January 2000	&	37.38	&	7	&	10:18:16.85	&	41:25:26.06	&	19.99	&	99	&	61	&		96	&		8	&		88	&	-0.837	\\
NGC\,3198	&	9551	&	5. February 2008	&	61.63	&	7	&	10:19:54.95	&	45:32:58.68	&	19.24	&	112	&	73	&		108	&		25	&		83	&	-0.535	\\
IC 2574\tablenotemark{*}	&	9541	&	30. June 2008	&	10.78	&	3	&	\nodata	&	\nodata	&	20.94	&	\nodata	&	6	&	$<$	2	&	$<$	1	&	$<$	1	&	\nodata	\\
Mrk 33	&	9519	&	1. February 2008	&	18.25	&	7	&	10:32:32.01	&	54:24:02.16	&	22.40	&	103	&	70	&		100	&		18	&		82	&	-0.640	\\
NGC\,3351\tablenotemark{*}	&	5931	&	1. July 2005	&	39.55	&	7	&	\nodata	&	\nodata	&	17.67	&	\nodata	&	572	&	$<$	33	&	$<$	6	&	$<$	32	&	\nodata	\\
NGC\,3521\tablenotemark{*}	&	9552	&	28. January 2008	&	69.69	&	3	&	11:05:48.51	&	-00:02:09.50	&	16.64	&	302	&	159	&		292	&		74	&		218	&	-0.492	\\
NGC\,3621	&	9278	&	6. March 2008	&	19.45	&	7	&	11:18:16.50	&	-32:48:50.59	&	20.00	&	23	&	39	&		21	&		3	&		19	&	-0.76	\\
NGC\,3627	&	9548	&	3. November 1999	&	49.54	&	7	&	11:20:15.00	&	12:59:30.40	&	15.95	&	266	&	234	&		251	&		57	&		195	&	-0.549	\\
NGC\,3938	&	7862	&	16. October 2008	&	4.89	&	7	&	\nodata	&	\nodata	&	19.82	&	\nodata	&	21	&	$<$	3	&	$<$	2	&	$<$	2	&	\nodata	\\
NGC\,4125	&	2071	&	9. September 2001	&	62.97	&	7	&	12:08:05.95	&	65:10:27.45	&	16.86	&	310	&	920	&		255	&		20	&		235	&	-0.841	\\
NGC\,4236	&	9543	&	20. February 2008	&	10.87	&	3	&	\nodata	&	\nodata	&	20.89	&	\nodata	&	9	&	$<$	2	&	$<$	2	&	$<$	1	&	\nodata	\\
NGC\,4254	&	7863	&	21. November 2007	&	5.07	&	7	&	\nodata	&	\nodata	&	18.47	&	\nodata	&	30	&	$<$	5	&	$<$	2	&	$<$	4	&	\nodata	\\
NGC\,4321\tablenotemark{*}	&	6727	&	18. February 2006	&	37.55	&	7	&	12:22:54.98	&	15:49:20.18	&	19.83	&	168	&	629	&		136	&		7	&		129	&	-0.898	\\
NGC\,4450	&	3997	&	30. April 2003	&	1.59	&	7	&	12:28:29.59	&	17:05:05.81	&	21.00	&	223	&	12	&		222	&		34	&		188	&	-0.695	\\
NGC\,4552	&	2072	&	22. April 2001	&	53.79	&	7	&	12:35:39.81	&	12:33:23.05	&	18.68	&	1856	&	3372	&		1675	&		157	&		1518	&	-0.812	\\
NGC\,4559\tablenotemark{*}	&	2027	&	4. June 2001	&	8.87	&	7	&	\nodata	&	\nodata	&	20.94	&	\nodata	&	38	&	$<$	5	&	$<$	2	&	$<$	5	&	\nodata	\\
NGC\,4569	&	5911	&	13. November 2005	&	31.29	&	7	&	12:36:49.81	&	13:09:46.09	&	20.45	&	681	&	259	&		668	&		84	&		584	&	-0.748	\\
NGC\,4579	&	807	&	2. May 2000	&	29.95	&	7	&	12:37:43.52	&	11:49:05.61	&	18.62	&	19026	&	1637	&		18938	&		4316	&		14622	&	-0.544	\\
NGC\,4594\tablenotemark{*}	&	9533	&	2. December 2008	&	79.74	&	3	&	12:39:59.47	&	-11:37:22.13	&	13.34	&	5154	&	454	&		5120	&		1460	&		3660	&	-0.430	\\
NGC\,4625	&	9549	&	5. March 2008	&	54.85	&	7	&	\nodata	&	\nodata	&	18.87	&	\nodata	&	76	&	$<$	10	&	$<$	7	&	$<$	8	&	\nodata	\\
NGC\,4631	&	797	&	16. April 2000	&	57.32	&	7	&	\nodata	&	\nodata	&	19.30	&	\nodata	&	97	&	$<$	10	&	$<$	7	&	$<$	7	&	\nodata	\\
NGC\,4725\tablenotemark{*}	&	2976	&	2. December 2002	&	24.64	&	7	&	12:50:26.59	&	25:30:02.82	&	21.00	&	257	&	114	&		252	&		7	&		244	&	-0.941	\\
NGC\,4736\tablenotemark{*}	&	808	&	13. May 2000	&	43.72	&	7	&	12:50:53.07	&	41:07:12.99	&	14.03	&	3851	&	1354	&		3754	&		428	&		3327	&	-0.772	\\
DDO 154	&	9544	&	3. April 2008	&	59.2	&	7	&	\nodata	&	\nodata	&	20.94	&	\nodata	&	64	&	$<$	6	&	$<$	4	&	$<$	5	&	\nodata	\\
NGC\,4826	&	411	&	27. March 2000	&	1.8	&	7	&	12:56:43.60	&	21:40:57.98	&	20.62	&	22	&	38	&		20	&		0	&		20	&	\nodata	\\
DDO 165	&	9537	&	8. February 2009	&	0	&	7	&	\nodata	&	\nodata	&	20.94	&	\nodata	&	12	&	$<$	2	&	$<$	1	&	$<$	2	&	\nodata	\\
NGC\,5033	&	412	&	28. April 2000	&	2.9	&	7	&	13:13:27.47	&	36:35:38.15	&	21.00	&	1148	&	74	&		1144	&		346	&		799	&	-0.396	\\
NGC\,5055	&	2197	&	27. August 2001	&	28	&	7	&	13:15:49.28	&	42:01:45.77	&	18.70	&	370	&	252	&		357	&		78	&		278	&	-0.562	\\
NGC\,5194\tablenotemark{*}	&	3932	&	7. August 2003	&	41.02	&	7	&	13:29:52.71	&	47:11:42.98	&	17.17	&	915	&	1462	&		830	&		76	&		754	&	-0.818	\\
NGC\,5195	&	414	&	23. January 2000	&	1.14	&	7	&	13:29:59.49	&	47:15:57.86	&	21.00	&	12	&	16	&		11	&		2	&		9	&	-0.64	\\
NGC\,5408\tablenotemark{*}	&	4558	&	29. January 2005	&	4.1	&	7	&	\nodata	&	\nodata	&	20.94	&	\nodata	&	14	&	$<$	2	&	$<$	1	&	$<$	2	&	\nodata	\\
NGC\,5474	&	9546	&	3. December 2007	&	29.76	&	7	&	\nodata	&	\nodata	&	20.94	&	\nodata	&	37	&	$<$	4	&	$<$	3	&	$<$	3	&	\nodata	\\
NGC\,5866 	&	2879	&	14. November 2002	&	30.26	&	7	&	15:06:29.58	&	55:45:48.48	&	18.51	&	40	&	242	&		27	&		19	&		8	&	0.389	\\
IC 4710\tablenotemark{*}	&	9570	&	25. June 2008	&	30.36	&	7	&	\nodata	&	\nodata	&	20.91	&	\nodata	&	34	&	$<$	6	&	$<$	5	&	$<$	3	&	\nodata	\\
NGC\,6822	&	2925	&	4. November 2002	&	20.21	&	3	&	\nodata	&	\nodata	&	21.25	&	\nodata	&	5	&	$<$	1	&	$<$	1	&	$<$	0	&	\nodata	\\
NGC\,6946\tablenotemark{*}	&	1043	&	7. September 2001	&	58.29	&	7	&	20:34:52.31	&	60:09:14.39	&	18.77	&	549	&	239	&		536	&		100	&		436	&	-0.628	\\
NGC\,7331	&	2198	&	27. January 2001	&	27.9	&	7	&	22:37:04.00	&	34:24:55.53	&	17.11	&	152	&	291	&		135	&		29	&		106	&	-0.577	\\
NGC\,7552	&	7848	&	31. March 2007	&	5.08	&	7	&	23:16:10.66	&	-42:35:03.64	&	21.00	&	207	&	238	&		196	&		13	&		183	&	-0.87	\\
NGC\,7793	&	3954	&	6. September 2003	&	46.41	&	7	&	\nodata	&	\nodata	&	20.94	&	\nodata	&	30	&	$<$	8	&	$<$	4	&	$<$	7	&	\nodata	\\
\enddata
\
\tablenotetext{a}{The ratio of the area of the background extraction region to the area of the source extraction region.}
\tablenotetext{*}{These galaxies have multiple $Chandra$ observations; we present here only the results from the longest and/or most recent data sets. }
\label{Table:tbl2}
\end{deluxetable} 

\begin{deluxetable}{lccr} 
\tabletypesize{\scriptsize}
\setlength{\tabcolsep}{0.04in}
\tablecaption{Nuclear Source Fluxes and Luminosities} 
\tablewidth{0pt} 
\tablehead{ 
\colhead{} & 
\colhead{0.3-8 keV} & 
\colhead{0.3-8 keV} & 
\colhead{} \\
\colhead{} & 
\colhead{Count rate} & 
\colhead{Flux} & 
\colhead{log $L_{0.3-8 \rm{keV}}$} \\
\colhead{Object} & 
\colhead{(10$^{-3}$ counts s$^{-1}$)} & 
\colhead{(10$^{-14}$ \ergscm)} & 
\colhead{(\ergs)} \\
} 
\startdata
NGC\,0024	&	$<$	2			&	$<$	0.12			&	$<$	36.97			\\
NGC\,0628	&		22	$\pm$	2	&		1.48	$\pm$	0.15	&		38.33	$\pm$	0.04	\\
NGC\,0855	&		18	$\pm$	2	&		1.18	$\pm$	0.12	&		38.21	$\pm$	0.03	\\
NGC\,0925	&		57	$\pm$	16	&		3.85	$\pm$	1.08	&		38.67	$\pm$	0.12	\\
NGC\,1097	&		3109	$\pm$	88	&		208.5	$\pm$	5.91	&		40.85	$\pm$	0.01	\\
NGC\,1291	&		289	$\pm$	9	&		19.38	$\pm$	0.59	&		39.34	$\pm$	0.01	\\
NGC\,1316	&		237	$\pm$	10	&		15.92	$\pm$	0.64	&		40.12	$\pm$	0.02	\\
NGC\,1404	&		183	$\pm$	6	&		20.57	$\pm$	0.73	&		40.19	$\pm$	0.02	\\
NGC\,1482	&		150	$\pm$	8	&		10.09	$\pm$	0.52	&		39.77	$\pm$	0.02	\\
NGC\,1705	&	$<$	2			&	$<$	0.13			&	$<$	36.72			\\
NGC\,2403	&	$<$	1			&	$<$	0.06			&	$<$	35.96			\\
M81 DwA	&	$<$	1			&	$<$	0.06			&	$<$	35.97			\\
DDO 053	&	$<$	2			&	$<$	0.11			&	$<$	36.19			\\
NGC\,2798	&		143	$\pm$	17	&		9.58	$\pm$	1.12	&		39.84	$\pm$	0.05	\\
NGC\,2841	&		82	$\pm$	6	&		5.49	$\pm$	0.37	&		38.8	$\pm$	0.03	\\
NGC\,2915	&	$<$	5			&	$<$	0.35			&	$<$	36.48			\\
Ho I	&	$<$	2			&	$<$	0.11			&	$<$	36.21			\\
NGC\,2976	&		6	$\pm$	3	&		0.42	$\pm$	0.17	&		36.79	$\pm$	0.17	\\
NGC\,3031	&		835	$\pm$	13	&		57.62	$\pm$	0.90	&		38.93	$\pm$	0.01	\\
NGC\,3034	&	$<$	28			&	$<$	1.89			&	$<$	37.44			\\
Ho IX	&	$<$	2			&	$<$	0.15			&	$<$	36.33			\\
M81 DwB	&	$<$	3			&	$<$	0.18			&	$<$	36.43			\\
NGC\,3190	&		111	$\pm$	7	&		7.46	$\pm$	0.50	&		39.43	$\pm$	0.03	\\
NGC\,3184	&		26	$\pm$	3	&		1.73	$\pm$	0.18	&		38.18	$\pm$	0.04	\\
NGC\,3198	&		18	$\pm$	2	&		1.18	$\pm$	0.11	&		38.13	$\pm$	0.04	\\
IC 2574	&	$<$	1			&	$<$	0.17			&	$<$	36.38			\\
Mrk 33	&		48	$\pm$	5	&		3.84	$\pm$	0.38	&		39.26	$\pm$	0.05	\\
NGC\,3351	&	$<$	8			&	$<$	0.55			&	$<$	37.76			\\
NGC\,3521	&		42	$\pm$	2	&		4.56	$\pm$	0.27	&		38.64	$\pm$	0.03	\\
NGC\,3621	&		11	$\pm$	2	&	$<$	0.81	$\pm$	0.18	&		37.57	$\pm$	0.09	\\
NGC\,3627	&		51	$\pm$	3	&		3.40	$\pm$	0.21	&		38.51	$\pm$	0.03	\\
NGC\,3938	&	$<$	6			&	$<$	0.42			&	$<$	37.88			\\
NGC\,4125	&		41	$\pm$	3	&		2.72	$\pm$	0.17	&		39.17	$\pm$	0.03	\\
NGC\,4236	&	$<$	2			&	$<$	0.20			&	$<$	36.47		0	\\
NGC\,4254	&	$<$	9			&	$<$	0.62			&	$<$	38.47			\\
NGC\,4321	&		36	$\pm$	3	&		2.43	$\pm$	0.21	&		39.07	$\pm$	0.04	\\
NGC\,4450	&		1395	$\pm$	94	&		93.51	$\pm$	6.27	&		40.65	$\pm$	0.03	\\
NGC\,4552	&		312	$\pm$	8	&		20.88	$\pm$	0.51	&		40	$\pm$	0.01	\\
NGC\,4559	&	$<$	6			&	$<$	0.38			&	$<$	37.79			\\
NGC\,4569	&		214	$\pm$	8	&		14.31	$\pm$	0.55	&		39.84	$\pm$	0.02	\\
NGC\,4579	&		6323	$\pm$	46	&		423.80	$\pm$	3.08	&		41.31	$\pm$	0	\\
NGC\,4594	&		642	$\pm$	9	&		69.87	$\pm$	0.98	&		40.2	$\pm$	0.01	\\
NGC\,4625	&	$<$	2			&	$<$	0.13			&	$<$	37.14			\\
NGC\,4631	&	$<$	2			&	$<$	0.12			&	$<$	37.05			\\
NGC\,4725	&		102	$\pm$	6	&		6.86	$\pm$	0.43	&		39.38	$\pm$	0.03	\\
NGC\,4736	&		859	$\pm$	14	&		57.66	$\pm$	0.94	&		39.29	$\pm$	0.01	\\
DDO 154	&	$<$	1			&	$<$	0.07			&	$<$	36.38			\\
NGC\,4826	&		112	$\pm$	25	&		8.34	$\pm$	1.86	&		38.5	$\pm$	0.1	\\
DDO 165	&	$<$	2			&	$<$	0.11			&	$<$	36.22			\\
NGC\,5033	&		3944	$\pm$	117	&		265.0	$\pm$	7.86	&		40.75	$\pm$	0.01	\\
NGC\,5055	&		127	$\pm$	7	&		8.55	$\pm$	0.45	&		38.84	$\pm$	0.02	\\
NGC\,5194	&		202	$\pm$	7	&		13.57	$\pm$	0.47	&		39.04	$\pm$	0.02	\\
NGC\,5195	&		99	$\pm$	29	&		7.37	$\pm$	2.20	&		38.38	$\pm$	0.13	\\
NGC\,5408	&	$<$	6			&	$<$	0.40			&	$<$	36.99			\\
NGC\,5474	&	$<$	1			&	$<$	0.09			&	$<$	36.71			\\
NGC\,5866 	&		9	$\pm$	2	&		0.60	$\pm$	0.11	&		38.05	$\pm$	0.08	\\
IC 4710	&	$<$	2			&	$<$	0.13			&	$<$	37.04			\\
NGC\,6822	&	$<$	1			&	$<$	0.08			&	$<$	34.51			\\
NGC\,6946	&		92	$\pm$	4	&		6.75	$\pm$	0.29	&		38.39	$\pm$	0.02	\\
NGC\,7331	&		48	$\pm$	4	&		3.28	$\pm$	0.28	&		38.99	$\pm$	0.04	\\
NGC\,7552	&		385	$\pm$	28	&		28.70	$\pm$	2.05	&		40.23	$\pm$	0.03	\\
NGC\,7793	&	$<$	2			&	$<$	0.11			&	$<$	36.13			\\
\enddata
\
\label{Table:tbl3}
\end{deluxetable} 

%

\begin{deluxetable}{lccccr} 
\tabletypesize{\scriptsize}
\setlength{\tabcolsep}{0.04in}
\tablecaption{Fit Parameters for IR/X-Ray Correlations} 
\tablewidth{0pt} 
\tablehead{ 
\colhead{Wavelength/} & 
\colhead{} & 
\colhead{} & 
\colhead{Spearman} &
\colhead{Pearson} &
\colhead{}\\
\colhead{Band} & 
\colhead{$L/L_{0.3-8 keV}$} & 
\colhead{Intercept}  &
\colhead{Coefficient} &
\colhead{Coefficient} &
\colhead{$P(r)$\tablenotemark{a}} \\} 
\startdata
3.6 $\mu$m      & 0.365		& 28.75	& 0.67 & 0.65 & $<$0.1\% \\
4.5 $\mu$m      & 0.360    	& 28.68	& 0.70 & 0.66 & $<$0.1\% \\
5.8 $\mu$m      & 0.295		& 31.28	& 0.61 & 0.59 & $<$0.1\% \\
8 $\mu$m        & 0.212		& 34.58	& 0.36 & 0.44 & $\sim$1\%   \\
24 $\mu$m       & 0.243 	& 32.94	& 0.28 & 0.38 & $\sim$10\%  \\ 
70 $\mu$m       & 0.218		& 34.62	& 0.27 & 0.27 & $\sim$10\%  \\
160 $\mu$m      & 0.249		& 33.52	& 0.27 & 0.27 & $\sim$10\%  \\
$J$ Band (global)     & 0.372	& 29.25 & 0.64 & 0.63 & $<$0.1\% \\
$H$ Band (global)     & 0.386	& 28.6  & 0.66 & 0.64 & $<$0.1\% \\
$K$ Band (global)     & 0.388	& 28.35 & 0.66 & 0.65 & $<$0.1\% \\
$J$ Band (nuclear)     & 0.643		& 17.83 & 0.77 & 0.77 & $<$0.1\% \\
$H$ Band (nuclear)     & 0.632		& 18.59	& 0.76 & 0.76 & $<$0.1\% \\
$K$ Band (nuclear)     & 0.661		& 17.58	& 0.78 & 0.74 & $<$0.1\% \\
\enddata

\tablenotetext{a}{Probability that there is in fact no correlation between the two parameters}
\label{Table:tbl5}
\end{deluxetable} 

\end{document}